\documentclass{jkas}

\def\year{2024} % publication year
\def\volume{57} % journal volume
\def\issue{2} % journal issue
\def\beginpage{1} % first page of article
\def\received{May 27, 2024} % date paper was received by JKAS
\def\accepted{July 30, 2024} % date of acceptance
\def\published{January 1, 2000} % date of publication
\date{Received \received; Accepted \accepted; Published \published}

\def\sun{\hbox{$\odot$}}

\def\arcsec{\hbox{$^{\prime\prime}$}}

\def\farcs{\hbox{$.\!\!^{\prime\prime}$}}

\def\smallplotfour#1#2#3#4{\centering \leavevmode
\includegraphics[width=89.5mm]{#1} \includegraphics[width=89.5mm]{#2}
\hfil \includegraphics[width=89.5mm]{#3} \includegraphics[width=89.5mm]{#4}}

\def\xsmallplotfour#1#2#3#4{\centering \leavevmode
\includegraphics[width=75mm]{#1} \includegraphics[width=75mm]{#2}
 \hfil \includegraphics[width=75mm]{#3} \includegraphics[width=75mm]{#4}}

\def\xsmallplotsix#1#2#3#4#5#6{\centering \leavevmode
\includegraphics[width=75mm]{#1} \includegraphics[width=75mm]{#2}
 \hfil \includegraphics[width=75mm]{#3} \includegraphics[width=75mm]{#4}
 \hfil \includegraphics[width=75mm]{#5} \includegraphics[width=75mm]{#6}}

\def\largeplottwo#1#2{\centering \leavevmode
\includegraphics[width=150mm]{#1}
 \hfil \includegraphics[width=150mm]{#2} }

\title{%
AGB and Post-AGB Stars versus Planetary Nebulae and Young Stellar Objects: Properties in Visual and IR Bands
}

\author[$\star$]{Kyung-Won Suh}{0000-0001-9104-9763}
\affil[1]{Department of Astronomy and Space Science, Chungbuk National University, 1, Chungdae-ro, Seowon-gu, Cheongju-si,
Chungcheongbuk-do 28644, Republic of Korea}
\def\corrauthor{%
K.-W. Suh, \email{kwsuh@chungbuk.ac.kr}
}

\def\runningauthor{%
Suh
}

\def\runningtitle{%
AGB and Post-AGB Stars versus Planetary Nebulae and YSOs
}

\def\keywords{%
stars: AGB and post-AGB --- circumstellar matter
--- dust, extinction --- infrared: stars --- stars: pre-main sequences
--- radiative transfer
}

\def\abstracttext{%
We investigate the properties of AGB and post-AGB (PAGB) stars, planetary 
nebulae, and young stellar objects (YSOs) in our Galaxy through an analysis of 
observational data covering visual and infrared (IR) wavelengths. Utilizing 
datasets from IRAS, 2MASS, AllWISE, Gaia DR3, and the SIMBAD object database, we 
perform an in-depth comparison between observational data and theoretical models. 
For this comparison, we present various color-magnitude diagrams (CMDs) in visual 
and IR bands, as well as IR two-color diagrams (2CDs). Our results demonstrate 
that the CMDs, incorporating the latest distance and extinction data from Gaia 
DR3 for a majority of sample stars, are effective in distinguishing different 
classes of stars. To improve the precision of our analysis, we apply theoretical 
radiative transfer models for dust shells around AGB and PAGB stars. A thorough 
comparison of the theoretical models with observations across various IR 2CDs and 
CMDs shows a significant agreement. We find that AGB and PAGB stars are among the 
brightest classes in visual and IR bands. Furthermore, most YSOs are clearly 
distinguishable from AGB stars on various IR CMDs, exhibiting fainter absolute 
magnitudes in IR bands. 
}

\begin{document}
\jkashead %% set title, authors, abstract, etc.

%%%%%%%%%%%%%%%%%%%%%%%%%%%%%%%%%%%%%%%%%%%%%%%%%%%%%%%%%%%%%%%%%%%%%
%%% BEGIN MAIN TEXT HERE %%%%%%%%%%%%%%%%%%%%%%%%%%%%%%%%%%%%%%%%%%%%
%%%%%%%%%%%%%%%%%%%%%%%%%%%%%%%%%%%%%%%%%%%%%%%%%%%%%%%%%%%%%%%%%%%%%

\section{Introduction\label{sec:intro}}

Asymptotic giant branch (AGB) stars are typically categorized as either O-rich
AGB (OAGB) or C-rich AGB (CAGB) stars, depending on the chemical composition of
their photosphere and/or outer envelope (e.g., \citealt{suh2021}). Almost all AGB
stars can be recognized as Long-Period Variables (LPVs), and the expansive
envelopes that surround these stars are highly conducive to substantial dust
formation (e.g., \citealt{hofner2018}).

As a star leaves the AGB phase, the mass loss reduces the remaining H-rich
envelope's mass, causing the stellar envelope to shrink and the effective
temperature to increase (e.g., \citealt{schoenberner1983}). This increase
continues until the central star is hot enough (around 30,000 K) to ionize the
circumstellar nebula, leading to the formation of a planetary nebula (PN).
Planetary nebulae (PNe) are considered a typical final stage in the stellar
evolution of many stars with masses ranging from 1 to 8 $M_{\odot}$ (e.g.,
\citealt{kwok2000}).

The transitional phase between the end of the AGB phase and the onset of the PN
phase is known as the post-AGB (PAGB) phase. In this period, the dust shell
produced during the AGB phase separates from the central star and becomes
optically thin within a few hundred years (e.g., \citealt{szczerba2007}).

Both young stellar objects (YSOs) and AGB stars show similar infrared (IR)
properties due to their low effective temperatures and cold, dense circumstellar
dust envelopes. Consequently, YSOs and AGB stars are commonly identified based on
their locations in IR two-color diagrams (2CDs) (e.g., \citealt{koenig2014};
\citealt{suh2021}). However, the IR colors of YSOs and AGB stars overlap
significantly in these diagrams, leading to potential misclassification (e.g.,
\citealt{lee2021}).

\begin{table*}
\centering
\caption{Sample stars in five classes (OAGB stars, CAGB stars, PAGB stars, PNe, and YSOs in our Galaxy) \label{tab:tab1}}
\begin{tabular}{lllllll}
\toprule
Class     & Reference & Number & Selected & IRAS & Gaia DR3 & AllWISE \\
\midrule
OAGB\_IC  &\citet{suh2022}  & 19,196 & 18,684$^{a}$ & 18,684  & 16,965  & 17,741  \\
OAGB\_NI  &\citet{suh2022}  & 45,413 & 44,864$^{a}$ & -       & 44,644  & 44,864  \\
OAGB  &\citet{suh2022}      & -      & 63,548       & 18,684  & 61,609  & 62,605  \\
\midrule
CAGB\_IC  &\citet{suh2024}  & 4909 & 4909 & 4909 & 4862  & 4752  \\
CAGB\_NI  &\citet{suh2024}  & 2254 & 2254 & -    & 2254  & 2252  \\
CAGB  &\citet{suh2024}      & -    & 7163 & 4909 & 7116  & 7004  \\
\midrule
PAGB  & SIMBAD$^{b}$  & 556  & 556  & 458  & 472  & 544    \\
\midrule
PNe(S)\_IC      &SIMBAD   & 1143  & 1143  & 1143  & 613 & 1143    \\
PNe   &\citet{gonzales2021} & 2035$^{c}$  & 2035$^{d}$  & - & 2035 & 1259  \\
\midrule
YSOs  &\citet{marton2023} & 11,671  & 11,537$^{d}$  & - & 11,537 & 9143  \\
\bottomrule
\end{tabular}
\begin{flushleft}
$^a$The list has been updated taking into account SIMBAD and other catalogs.
$^b$PAGB catalogs of \citet{szczerba2007} and \citet{kohoutek2001} are also considered.
$^c$Central stars of PNe (CSPNe).
$^d$Gaia DR3 sources.
\end{flushleft}
\end{table*}

\begin{table}
\scriptsize
\caption{Visual and IR bands and zero magnitude flux values\label{tab:tab2}}
\centering
\begin{tabular}{lllll}
\hline \hline
Band &$\lambda_{ref}$ ($\mu$m)	&ZMF (Jy) &Telescope &Reference 	\\
\hline
Bp      &0.511	&	3552	&	Gaia	& \citet{rimoldini2023}\\
G   	&0.622	&	3229	&	Gaia    & \citet{rimoldini2023}	\\
Rp   	&0.777	&	2555	&	Gaia    & \citet{rimoldini2023}	\\
K[2.2]  &2.159	&	666.7	&	2MASS	& \citet{cohen2003}\\
W2[4.6]	&4.60	&	170.663	&	WISE    & \citet{jarrett2011}	\\
IR[12]	&12     &  28.3 	&	IRAS     & \citet{beichman1988}	\\
W3[12]$^a$	&12	    &  28.3 	&	WISE & \citet{jarrett2011}\\
W4[22]	&22.08	&	8.284	&	WISE    & \citet{jarrett2011}	\\
IR[25]	&25 & 6.73 	&	IRAS      & \citet{beichman1988}	\\
IR[60]	&60 & 1.19 	&	IRAS      & \citet{beichman1988}	\\
\hline
\end{tabular}
\begin{flushleft}
$^a$See section 4.2 in ~\citealt{suh2020}).
\end{flushleft}
\end{table}

The Gaia Data Release 3 (DR3) has delivered valuable visual band data for over 
one billion stars (\citealt{rimoldini2023}). The newly acquired distance 
(\citealt{bailer-jones2021}) and extinction (e.g., \citealt{lallement2022}) 
information derived from Gaia DR3 is instrumental in determining the absolute 
magnitudes of AGB and PAGB stars, PNe, and YSOs within our Galaxy. 

In this study, we explore properties of AGB and PAGB stars, PNe, and YSOs within 
our Galaxy through a thorough analysis of observational data spanning visual and 
IR bands. In Section ~\ref{sec:sample}, we furnish lists of known sample stars. 
For the sample stars, we conduct cross-identifications with counterparts from 
IRAS, 2MASS, AllWISE, and Gaia DR3. 

Section~\ref{sec:visual-pro} presents color-magnitude diagrams (CMDs) using the 
Gaia DR3 data for distinct classes of sample stars. Section~\ref{sec:models} 
outlines the theoretical radiative transfer models for dust shells around AGB and 
PAGB stars. 

Section~\ref{sec:irpro} presents a variety of IR 2CDs and IR CMDs for different 
classes of sample stars, alongside theoretical models of AGB and PAGB stars. This 
comparison seeks to reveal potential differences in their IR properties. Finally, 
Section~\ref{sec:sum} compiles and summarizes the main findings and results of 
this paper.

\section{Sample Stars in our Galaxy \label{sec:sample}}

We have compiled lists of known OAGB stars, CAGB stars, PAGB stars, PNe, and YSOs 
in our Galaxy utilizing various literature sources and the SIMBAD astronomical 
database from the Strasbourg Astronomical Data Centre (CDS), as detailed in 
Table~\ref{tab:tab1}. For each class, the numbers of counterparts in IRAS, Gaia 
DR3, and AllWISE are listed.

\subsection{AGB and Post-AGB stars\label{sec:agb}}

\citet{suh2021}, \citet{suh2022}, and \citet{suh2024} presented catalogs of AGB
stars in our Galaxy, divided into two categories: one with IRAS counterparts
(AGB\_IC objects) based on the IRAS source catalog for brighter or more isolated
stars, and another without IRAS counterparts (AGB\_NI objects) based on the
AllWISE or Gaia DR3 source catalog for fainter stars or those in crowded regions.
Most AGB\_IC objects have corresponding entries in the AllWISE or Gaia DR3
catalogs.

As detailed in Table~\ref{tab:tab1}, we use the list of OAGB stars from
\citet{suh2022}, updated with information from SIMBAD and a recent catalog of
carbon stars by \citet{suh2024}.

OH/IR stars are typically regarded as more massive OAGB stars with denser dust 
envelopes and higher mass-loss rates (e.g., \citealt{suh2021}). From the sample 
of known OH/IR stars in SIMBAD, we select 202 extreme OH/IR (EOH/IR) stars, which 
exhibit the brightest absolute magnitudes in the W4[22] band. All of these 202 
EOH/IR stars belong to the OAGB\_IC class. The EOH/IR stars, a subclass of OAGB 
stars, can be considered the most massive (or evolved) AGB stars. 

For CAGB stars, we use the CAGB\_IC and CAGB\_NI objects from the catalog of CAGB
stars by \citet{suh2024}.

For PAGB stars, we considered 307 entries from the SIMBAD database, 326 objects 
from the PAGB catalog by \citet{szczerba2007}, and 334 pre-PNe (PPNe) from the 
Galactic PNe catalog by \citet{kohoutek2001}. After removing duplicates, we 
compiled a list of 556 unique objects for our sample of PAGB stars.

\subsection{Planetary Nebulae \label{sec:pne}}

\citet{gonzales2021} investigated the properties of central stars of PNe (CSPNe) 
using Gaia EDR3 data and presented a list of 2035 PNe in our Galaxy. This can be 
regarded as one of the most comprehensive and reliable catalogs of PNe in our 
Galaxy. 

We use two different groups for the list of PNe (see Table~\ref{tab:tab1}): a 
list of 1143 objects (PNe(S)\_IC) from the SIMBAD database with IRAS 
counterparts, and the catalog of 2035 PNe from \citet{gonzales2021}. A 
significant number of objects in these two groups appear to be duplicates. 
However, finding accurate counterparts is challenging due to the large beam size 
of IRAS, so we did not identify the IRAS counterparts for the PNe catalog from 
\citet{gonzales2021}. 

For constructing 2CDs or CMDs without using IRAS data, we use the PNe catalog 
from \citet{gonzales2021}. Conversely, for constructing IR 2CDs or CMDs using 
IRAS data, we use PNe(S)\_IC.

\subsection{Young Stellar Objects \label{sec:yso}}

\citet{marton2023} presented the Konkoly Optical YSO catalog containing 11,671 
objects, one of the most reliable and extensive catalogs of YSOs in our Galaxy. 
For this study, we use the 11,537 objects with Gaia DR3 counterparts listed in 
the original catalog as the sample stars in the class of YSOs (see 
Table~\ref{tab:tab1}). 

Identifying the proper IRAS counterpart for the YSOs is challenging because they 
are generally dimmer and/or smaller than most IRAS sources. Therefore, we use the 
YSO catalog for constructing 2CDs or CMDs without using IRAS data. 

There are various subclasses of YSOs. In this work, we consider three subclasses
identified in the Konkoly Optical YSO catalog (\citealt{marton2023}): (1) CTT
(3822 objects) - classical T Tauri stars, which are K–M type stars with emission
spectra and accretion disks; (2) WTT (4151 objects) - weak-line T Tauri stars,
identified by their G–M spectral types and strong Li I absorption lines; (3)
HAeBe (270 objects) - Herbig Ae/Be stars, which are YSOs in the higher mass range
(2 - 8 $M_{\odot}$).

\subsection{Photometric Data \label{sec:photdata}}

Table~\ref{tab:tab2} provides a list of the visual and IR bands employed in this 
study for 2CDs and CMDs. For each band, the reference wavelength 
($\lambda_{ref}$) and zero magnitude flux (ZMF) value are included, aiding in the 
derivation of theoretical model colors and magnitudes from the spectral energy 
distributions (SEDs) (refer to Section~\ref{sec:models}). For more details about 
the IR photometric data, see Section 2.1 in \citet{suh2021}. 

In this study, we exclusively utilize good-quality observational data across all
wavelength bands for the IR photometric data (with quality better than 1 for the
IRAS data and better than U for the WISE data).

\subsection{Cross-identification\label{sec:cross}}

The IRAS Point Source Catalog (PSC) contains 245,889 sources, and the IRAS Faint 
Source Catalog (FSC) includes 173,044 sources. In this study, we utilize the 
combined IRAS source catalog as presented by \citet{abrahamyan2015} and 
\citet{suh2024}, which consists of 345,164 sources, with 73,769 associated 
sources from the PSC and FSC. For the combined 345,164 IRAS sources, the 
counterparts from the AllWISE, 2MASS, and Gaia DR3 catalogs were obtained as 
explained in Section 2.3 of \citet{suh2024}.  

For the objects with IRAS counterparts (AGB\_IC, PAGB\_IC, and PNe(S)\_IC 
objects), we use the AllWISE, 2MASS, and Gaia DR3 counterparts obtained from the 
combined IRAS source catalog, which includes all available counterparts.

The AGB and PAGB stars without IRAS counterparts (AGB\_NI objects and PAGB stars 
without IRAS counterparts) were identified using AllWISE and/or Gaia DR3 sources. 
For objects without Gaia DR3 counterparts, we identified the nearest 2MASS 
counterparts within 3$\arcsec$ using the positions of the AllWISE sources. 

The catalogs of YSOs and PNe are entirely based on Gaia DR3 data, unlike the 
catalogs for AGB and PAGB stars and PNe(S)\_IC objects. For YSOs and PNe, we did 
not identify IRAS counterparts (see Table~\ref{tab:tab1}). 

For the objects with Gaia DR3 counterparts, we find the nearest AllWISE and 2MASS 
counterparts within 3$\arcsec$ using the positions of the Gaia DR3 sources. Note 
that the field-of-view pixel sizes for 2MASS and AllWISE are 2$\arcsec$ and 
2$\farcs$75, respectively.

\subsection{Distances and Galactic extinction \label{sec:ext}}

The recently acquired distance measurements (\citealt{bailer-jones2021}) and 
extinction data (\citealt{lallement2022}), derived from Gaia DR3 data, are 
crucial for calculating the absolute luminosity of the sample stars in our 
Galaxy. 

For a majority of sample stars, distances derived from Gaia DR3 data are readily 
accessible (see Table~\ref{tab:tab1}). Additionally, for CAGB stars, we employ 
distances obtained through a comparative analysis between the observed SED and 
the theoretical CAGB model SED in cases where Gaia DR3 distances are unavailable 
(see \citealt{suh2024}). 

After determining the distance and position of each object, we acquired 
extinction values ($A_V$) in the visual band from the data server provided by 
\citet{lallement2022}. Using the $A_V$ data, we can compute extinction values in 
Gaia and 2MASS bands, utilizing the optical to mid-infrared (MIR) extinction law 
outlined by \citet{wang2019}. 

In this work, Galactic extinction was considered only for visual and NIR bands, 
up to the K[2.2] band (see Table~\ref{tab:tab2}), because Galactic extinction is 
minor and more complicated at longer wavelengths. 

The acquired distance and extinction data are utilized in constructing various
2CDs and CMDs, as elaborated in Sections \ref{sec:visual-pro} and
\ref{sec:irpro}. The extinction data are incorporated into the 2CDs and CMDs
using the Gaia and/or 2MASS data outlined in these sections.

%%% FIGURE %%%%%%%%%%%%%%%%%%%%%%%%%%%%%%%%%%%%%%%%%%%%%%%%%%%%%%%%%%
\begin{figure*}
\centering
\includegraphics[width=150mm]{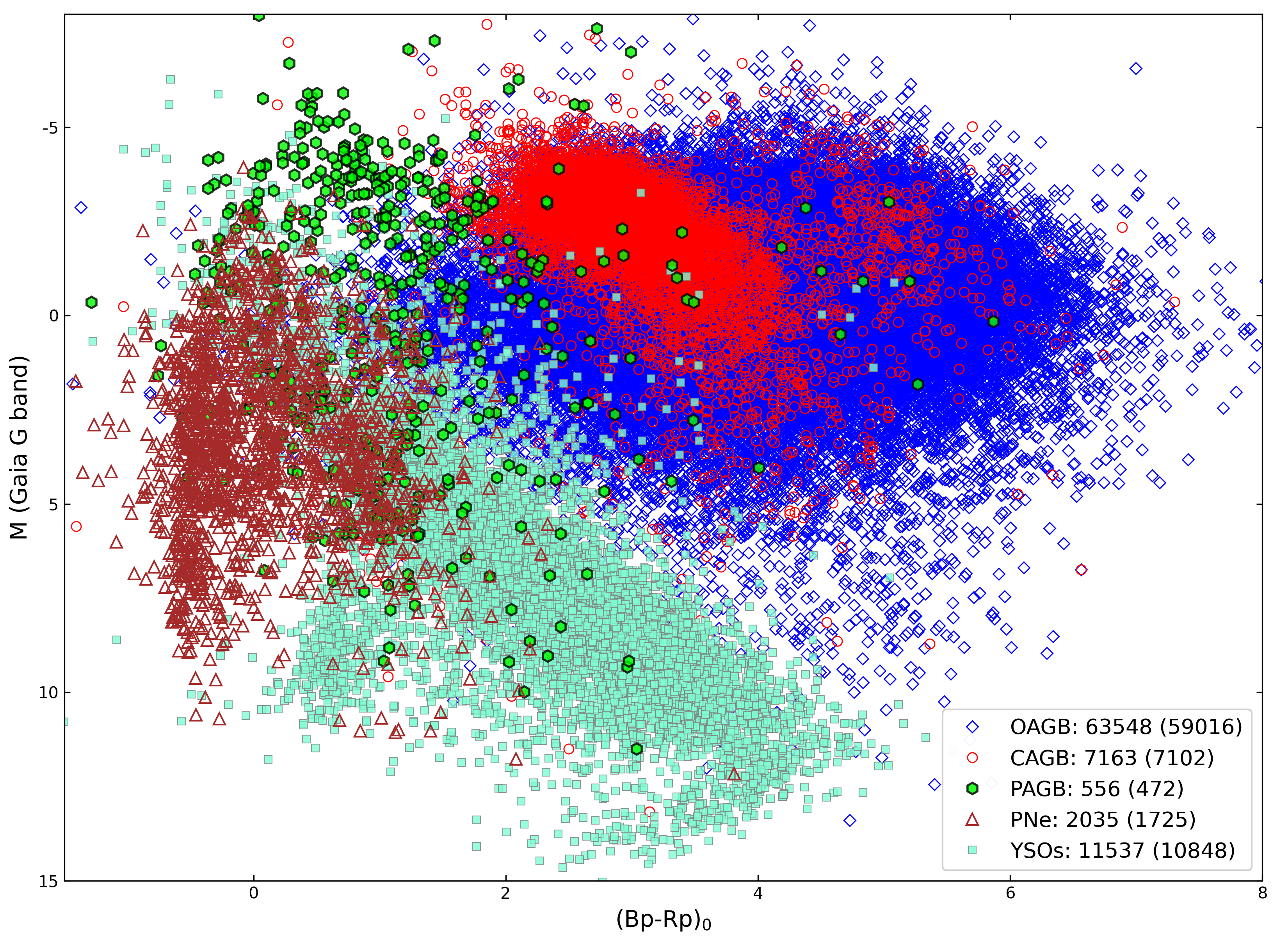}
%\caption{An example figure, from \citet{park2012}.\label{fig:jkasfig1}}
\caption{A Gaia DR3 CMD for the five classes of sample stars (see Table~\ref{tab:tab1}).
For each class, the number of objects is indicated.
The number in parentheses represents the count of plotted objects with good-quality observational data.
\label{f1}}
\end{figure*}
%%%%%%%%%%%%%%%%%%%%%%%%%%%%%%%%%%%%%%%%%%%%%%%%%%%%%%%%%%%%%%%%%%%%%

%%% FIGURE %%%%%%%%%%%%%%%%%%%%%%%%%%%%%%%%%%%%%%%%%%%%%%%%%%%%%%%%%%
\begin{figure}[t]
\centering
\includegraphics[width=88mm]{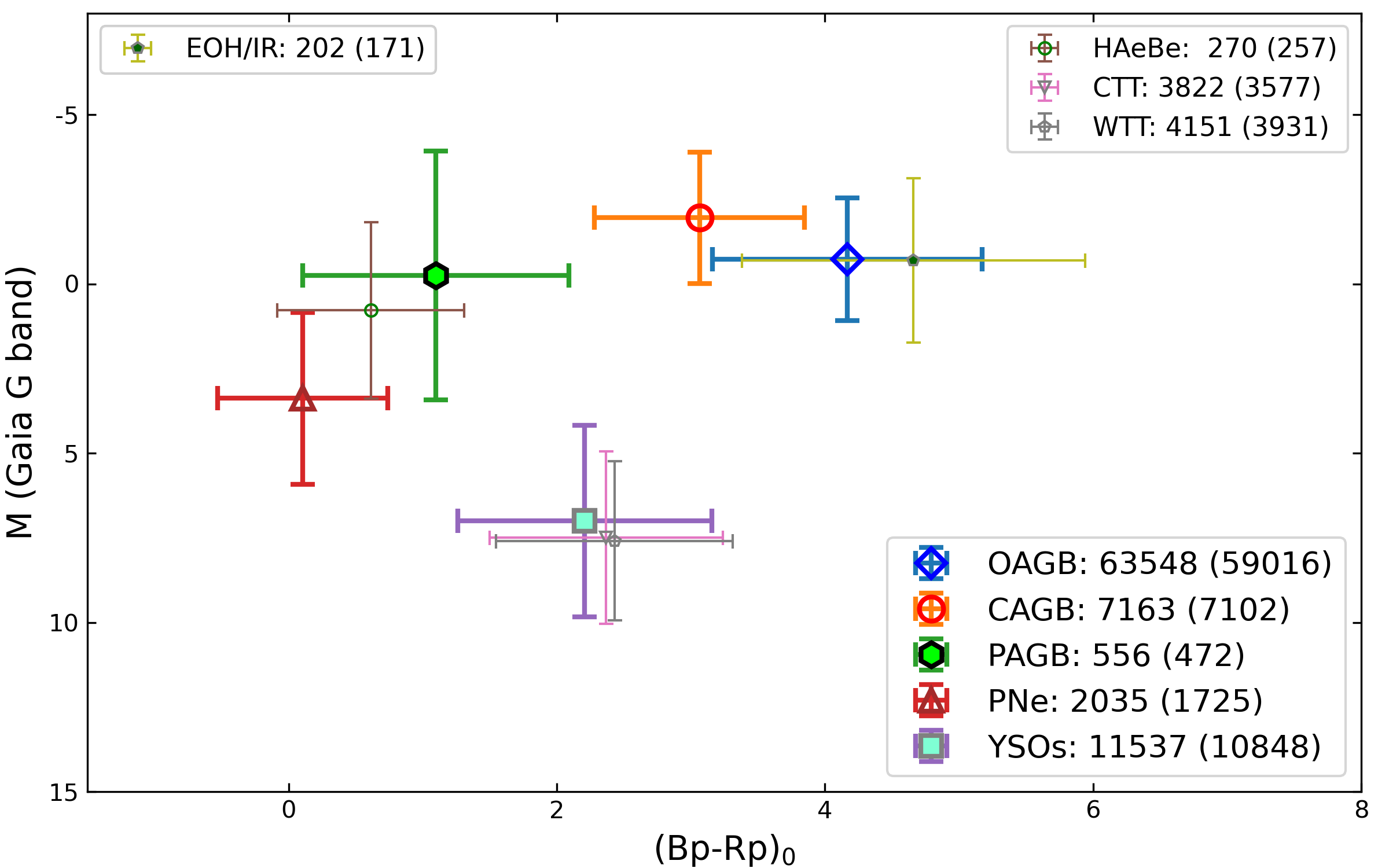}
\caption{An error-bar plot for the five classes of sample stars (see Figure~\ref{f1}).
Additionally, plots for three subclasses of YSO and one subclass (EOH/IR) of OAGB are shown.
\label{f2}}
%\vspace{5mm} %% add extra space ONLY when figures/tables are "colliding"!
\end{figure}
%%%%%%%%%%%%%%%%%%%%%%%%%%%%%%%%%%%%%%%%%%%%%%%%%%%%%%%%%%%%%%%%%%%%%

%%% FIGURE %%%%%%%%%%%%%%%%%%%%%%%%%%%%%%%%%%%%%%%%%%%%%%%%%%%%%%%%%%
\begin{figure*}
\centering
\includegraphics[width=150mm]{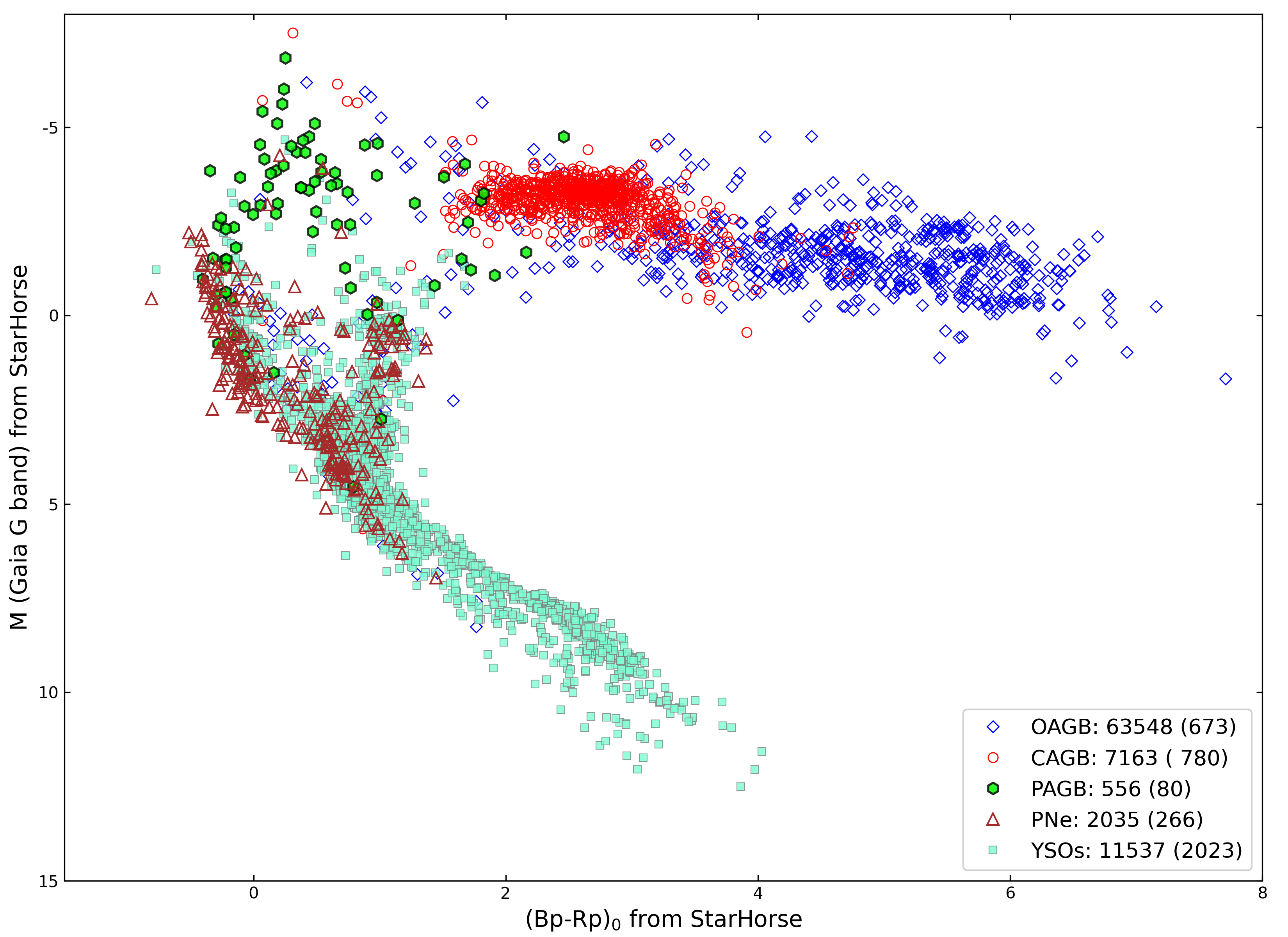}
%\caption{An example figure, from \citet{park2012}.\label{fig:jkasfig1}}
\caption{A Gaia DR3 CMD using StarHorse (\citealt{anders2022}) for the five classes of sample stars
(see Table~\ref{tab:tab1}).
For each class, the number of objects is indicated.
The number in parentheses represents the count of plotted objects with good-quality observational data.
\label{f3}}
\end{figure*}
%%%%%%%%%%%%%%%%%%%%%%%%%%%%%%%%%%%%%%%%%%%%%%%%%%%%%%%%%%%%%%%%%%%%%

%%% FIGURE %%%%%%%%%%%%%%%%%%%%%%%%%%%%%%%%%%%%%%%%%%%%%%%%%%%%%%%%%%
\begin{figure}[t]
\centering
\includegraphics[width=88mm]{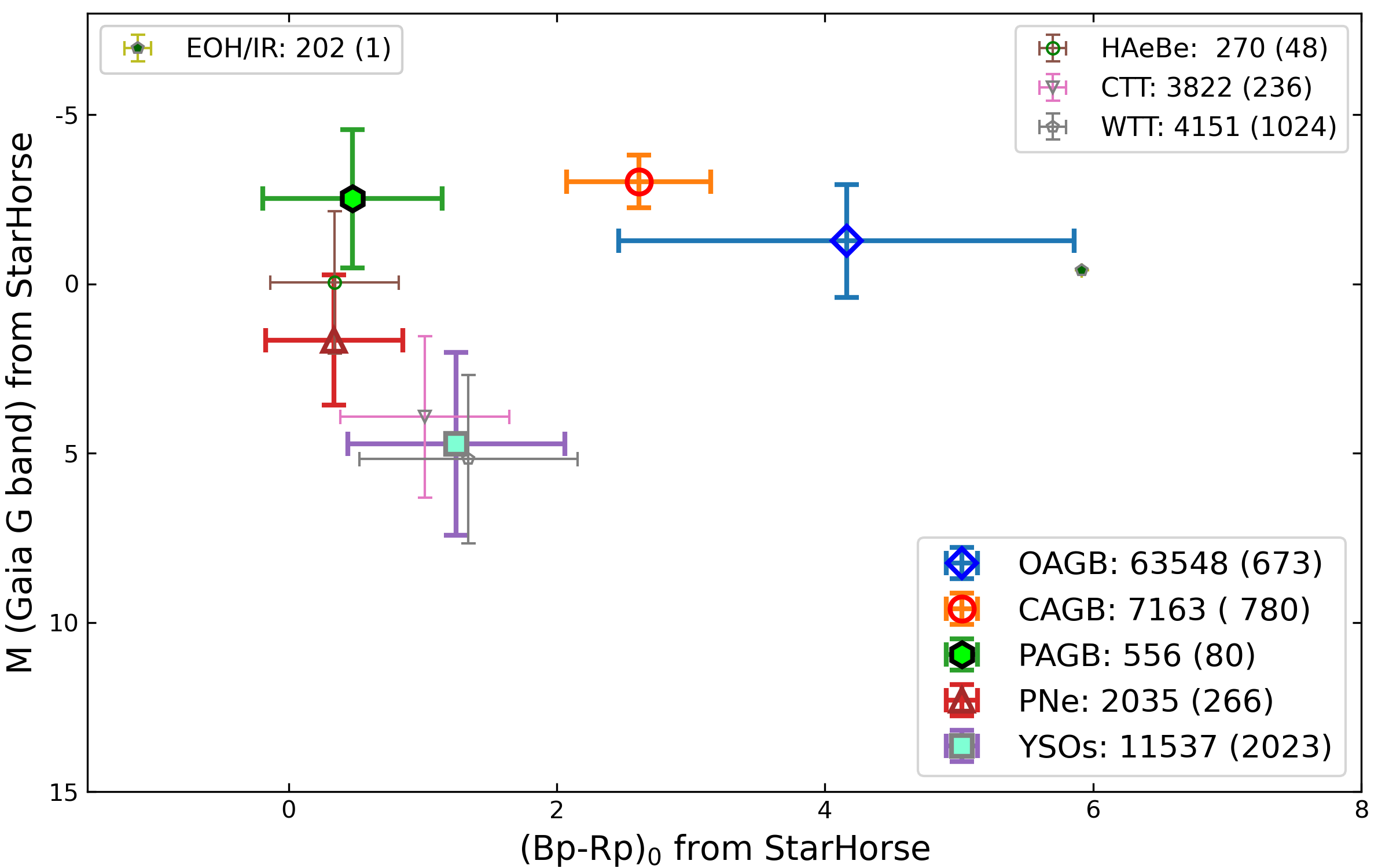}
\caption{An error-bar plot for the Gaia DR3 CMD using StarHorse (\citealt{anders2022}) for the five classes of sample stars (see Figure~\ref{f3}).
Additionally, plots for three subclasses of YSO and one subclass (EOH/IR) of OAGB are shown.
\label{f4}}
%\vspace{5mm} %% add extra space ONLY when figures/tables are "colliding"!
\end{figure}
%%%%%%%%%%%%%%%%%%%%%%%%%%%%%%%%%%%%%%%%%%%%%%%%%%%%%%%%%%%%%%%%%%%%%

\section{Properties in Visual bands\label{sec:visual-pro}}

In this section, we present a useful CMD using Gaia DR3 data: absolute magnitudes 
in the G band versus unreddened (Bp-Rp) colors. This CMD incorporates distance 
information from \citet{bailer-jones2021} and extinction information from 
\citet{lallement2022}, both derived from Gaia DR3 data. 

Figure~\ref{f1} shows the Gaia DR3 CMD for OAGB stars, CAGB stars, PAGB stars, 
PNe, and YSOs (see Table~\ref{tab:tab1}). Despite large scatter, the objects in 
these five classes are distributed in specific regions on the CMD. 
Figure~\ref{f2} shows an error-bar plot of the CMD. 

AGB stars are located in the upper-right region, PAGB stars in the upper-left
region, PNe in the middle-left region, and YSOs in the lower-middle region. Among
AGB stars, CAGB stars are more concentrated in the upper-middle region, while
OAGB stars are more scattered in the upper-right and upper-left regions.

For a selection of Gaia DR3 sources, \citet{anders2022} presented a catalog of 
362 million stellar parameters, distances, and extinctions based on Gaia EDR3, 
Pan-STARRS1, SkyMapper, 2MASS, and AllWISE. The new data and computational 
updates in their StarHorse code substantially improved the accuracy and precision 
over previous photo-astrometric stellar parameter estimates. We use the absolute 
magnitudes in the G band and unreddened (Bp-Rp) colors from their catalog, which 
are more reliable. 

Figure~\ref{f3} shows the Gaia DR3 CMD using StarHorse (\citealt{anders2022}) for 
OAGB stars, CAGB stars, PAGB stars, PNe, and YSOs (see Table~\ref{tab:tab1}). 
Although the number of plotted objects is smaller, the more reliable data points 
show more distinct separations among the classes of sample stars.  However, the 
general trends are similar to those in Figure~\ref{f1}. Figure~\ref{f4} shows an 
error-bar plot of the CMD. 

We find that AGB and PAGB stars are among the brightest classes in visual bands.
Generally, AGB and PAGB stars are brighter than PNe, and PNe are brighter than
YSOs in visual bands. However, HAeBe stars, a subclass of YSOs, show comparable
brightness to PAGB stars or PNe. The less massive CTT and WTT stars are located
in the lower region (see Figures~\ref{f2} and \ref{f4}).

For EOH/IR stars, which are the brightest subclass among the AGB classes in IR 
bands (see Section \ref{sec:irpro}), the absolute magnitudes in visual bands can 
be dimmer than those of other subclasses. This is because the thick dust 
envelopes surrounding EOH/IR stars cause greater extinction of emission in visual 
bands.

\begin{table}
%\scriptsize
\caption{Models for typical AGB stars\label{tab:tab3}}
\centering
\begin{tabular}{llllll}
\hline \hline
Model &Class &Dust$^1$ &$\tau_{10}$ & $T_*$ (K) & $L_* (10^{3} L_{\odot}$) \\
\hline
LM1  & OAGB  &silicate  &0.001 & 3000  & 1 \\
LM2  & OAGB  &silicate  &0.01 & 3000  & 2 \\
LM3  & OAGB  &silicate  &0.1  & 3000  & 3 \\
\hline
CA1  & CAGB  &AMC       &0.0001  & 3000  & 5 \\
CA2  & CAGB  &AMC       &0.01  & 2700  & 5 \\
CA3  & CAGB  &AMC       &0.1  & 2500  & 5 \\
CA4  & CAGB  &AMC       &0.5  & 2500  & 7 \\
CA5  & CAGB  &AMC       &1    & 2200  & 8 \\
CA6  & CAGB  &AMC       &2    & 2000  & 10 \\
CA7  & CAGB  &AMC       &4    & 2000  & 10 \\
\hline
HM4  & OAGB  &silicate  &7    & 2000  & 10 \\
HM5  & OAGB  &silicate  &15   & 2000  & 20 \\
HM6  & OAGB  &silicate  &30   & 2000  & 20 \\
\hline
\end{tabular}
\begin{flushleft}
\scriptsize
$^1$See Section~\ref{sec:agb-model} for details. For all models, $T_c$ = 1000 K.
\end{flushleft}
\end{table}

\section{Theoretical Dust Shell Models\label{sec:models}}

To investigate IR properties of AGB and post-AGB stars, the theoretical models
for dust shells are very useful because the objects have substantial dust
envelopes.

To calculate theoretical model SEDs for AGB and PAGB stars, we utilize radiative
transfer models designed for spherically symmetric dust shells surrounding
central stars. We employ the radiative transfer code RADMC-3D
(\url{http://www.ita.uni-heidelberg.de/~dullemond/software/radmc-3d/}), applying
the same methodologies as employed by \citet{sk2013}, \citet{suh2015}, and
\citet{suh2021}. For further insights into the theoretical models, refer to
\citet{suh2020}.

\subsection{Models for AGB stars\label{sec:agb-model}}

For AGB stars, we use the same models as employed by \citet{suh2024}. 
Table~\ref{tab:tab3} lists the parameters for six models representing typical 
OAGB stars and seven models representing typical CAGB stars. We assume a 
continuous power-law dust density distribution ($\rho \propto r^{-2}$). The inner 
radius of the dust shell is determined by the dust formation temperature ($T_c$), 
which is set at 1000 K. The outer radius is defined where the dust temperature 
reaches 30 K. We use spherical dust grains with a uniform radius of 0.1 $\mu$m. 
The dust optical depth ($\tau_{10}$) is measured at a reference wavelength of 10 
$\mu$m. We assume a stellar blackbody temperature ($T_*$) ranging from 2000 to 
3000 K and a stellar luminosity ($L_*$) ranging from $1 \times 10^{3}$ to $2 
\times 10^{4}$ $L_{\odot}$. 

For OAGB stars, we use the optical constants of silicate dust from
\citet{suh1999}. Specifically, warm silicate is used for OAGB stars with thin
dust shells (three models with $\tau_{10} \leq 3$), and cold silicate is used for
OAGB stars with thick dust shells (three models with $\tau_{10} > 3$). For CAGB
stars, we apply the optical constants of amorphous carbon (AMC) from
\citet{suh2000}.

\subsection{Models for post-AGB stars\label{sec:pagb-model}}

\citet{suh2015} introduced theoretical models depicting the spectral evolution of 
PAGB stars. The author employed radiative transfer models for the dust shells 
detaching from evolving central stars in the PAGB phase, utilizing parameters 
derived from the central stars as presented by \citet{vanHoof1997}. Various dust 
shell models with different mass-loss rates were utilized for each core-mass 
model of the central star to reproduce a variety of observations of PAGB stars. 
\citet{suh2018} investigated PAGB model tracks on IR 2CDs using WISE data for 
different samples of AGB stars, PAGB stars, and PNe.

In this study, we utilize the PAGB models outlined in \citet{suh2015}, modified
to suit the new IR CMDs and IR 2CDs. For the central star of a PAGB star, we
adopt pertinent model parameters from the theoretical frameworks presented by
\citet{vanHoof1997}. These models presume a constant stellar luminosity alongside
varying stellar blackbody temperatures throughout the PAGB phase. By the
conclusion of the PAGB phase, the stellar blackbody temperature reaches 30,000 K.

The initial dust optical depth ($\tau_{i}$) is defined as the dust optical depth
($\tau_{10}$) at the start of the PAGB phase. The expansion velocity
($v_{\text{exp}}$) and mass-loss rate ($\dot{M}$) of the dust shell are presumed
to remain constant throughout the PAGB phase. We adopt an expansion velocity of
$v_{\text{exp}}$ = 15 km s$^{-1}$.

For intermediate-mass O-rich ($\tau_{i}$=15) and C-rich ($\tau_{i}$=2) PAGB
stars, we use $M_{core}$ = 0.605 $M_{\sun}$ and $L_{*}$ = 6310 $L_{\sun}$. We use
the PAGB mass-loss rates (PMLs) at 1 and 5 times the standard value, for which
the PAGB time scales are 2093 and 832 years, respectively.

For high-mass O-rich stars ($\tau_{i}$=50), we use $M_{core}$ = 0.696 $M_{\sun}$
and $L_{*}$ = 11610 $L_{\sun}$ with PML at the standard value, for which the PAGB
time scale is 174 years.

\begin{figure*}
\centering
\largeplottwo{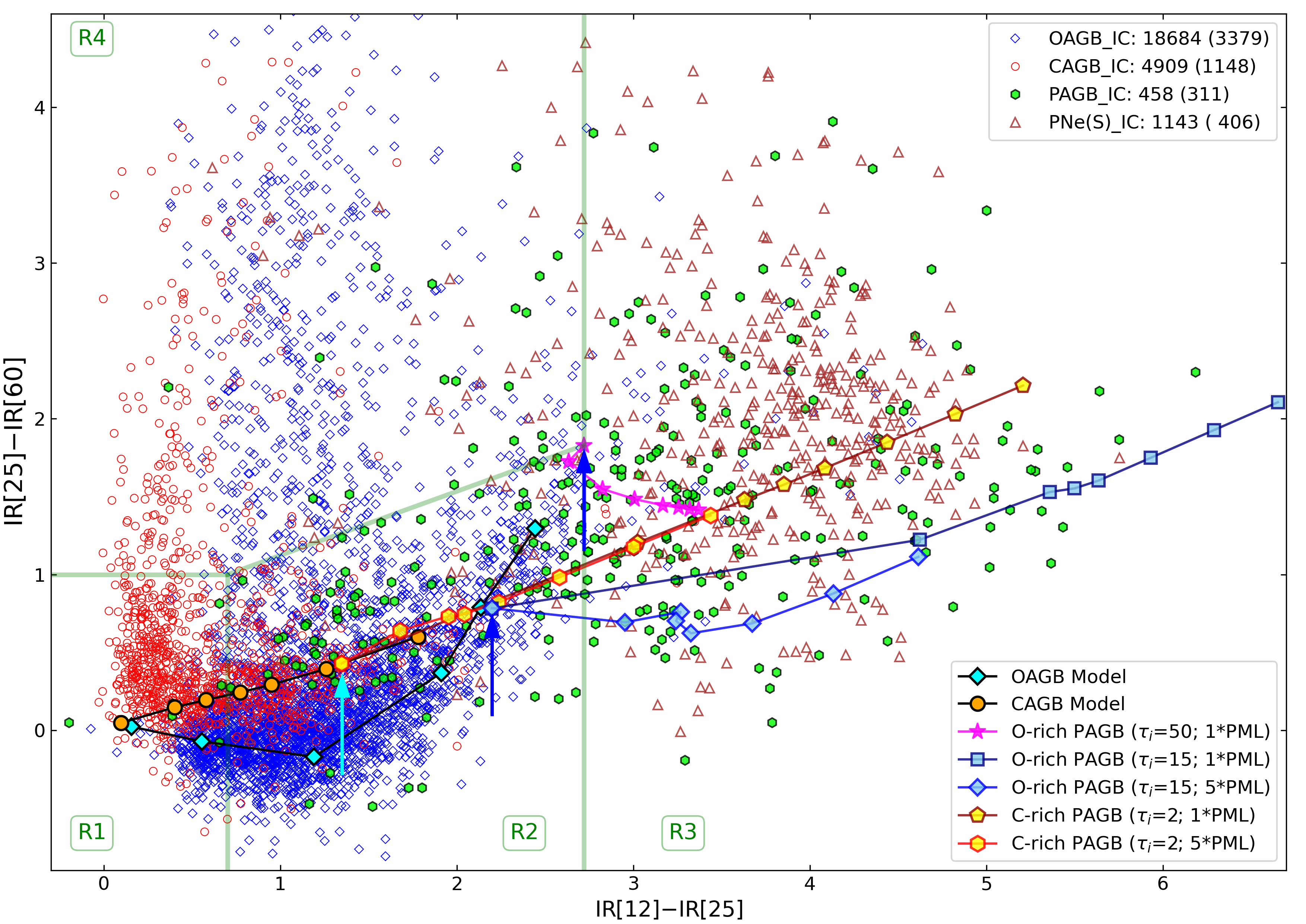}{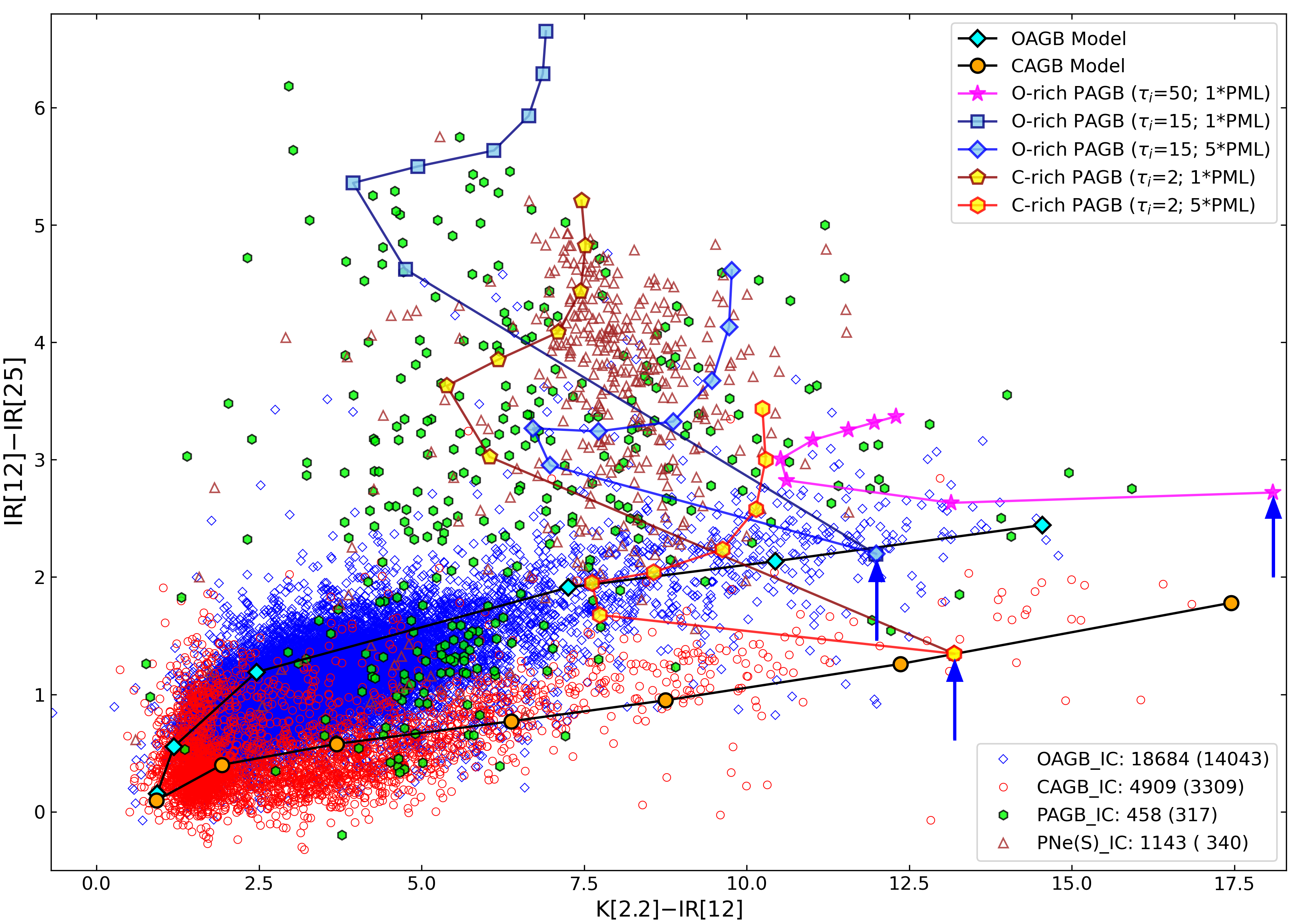}\caption{IRAS-2MASS 2CDs for AGB and PAGB stars and PNe with IRAS counterparts (see Table~\ref{tab:tab1}).
For CAGB models: $\tau_{10}$ = 0.0001, 0.01, 0.1, 0.5, 1, 2, and 4 from left to right.
For OAGB models: $\tau_{10}$ = 0.001, 0.01, 0.1, 7, 15, and 30 from left to right (see Section~\ref{sec:agb-model}).
For each class, the number of objects is indicated.
The number in parentheses represents the count of plotted objects with good-quality observational data.
The blue (or cyan) arrows indicate the start points of the PAGB models with various $\tau_{i}$ and PML (see Section~\ref{sec:pagb-model}).
\label{f5}}
\end{figure*}

\begin{figure*}
\centering
\includegraphics[width=150mm]{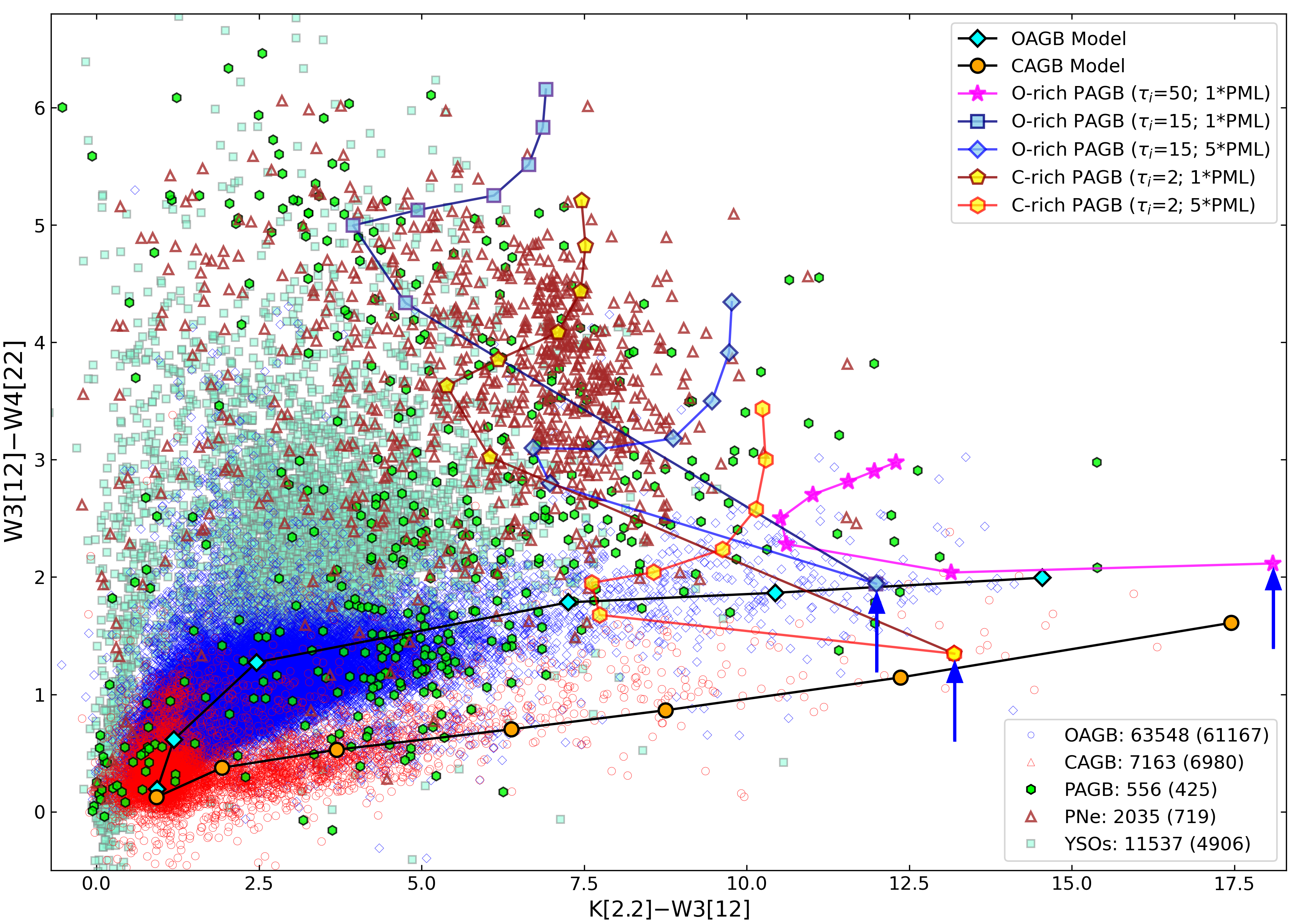}
%\caption{An example figure, from \citet{park2012}.\label{fig:jkasfig1}}
\caption{A WISE-2MASS 2CD for the five classes of sample stars (see Table~\ref{tab:tab1}).
For CAGB models: $\tau_{10}$ = 0.0001, 0.01, 0.1, 0.5, 1, 2, and 4 from left to right.
For OAGB models: $\tau_{10}$ = 0.001, 0.01, 0.1, 7, 15, and 30 from left to right (see Section~\ref{sec:agb-model}).
For each class, the number of objects is indicated.
The number in parentheses represents the count of plotted objects with good-quality observational data.
The blue arrows indicate the start points of the PAGB models with various $\tau_{i}$ and PML models (1*PML and 5*PML) (see Section~\ref{sec:pagb-model}).
\label{f6}}
\end{figure*}

\begin{figure*}
\centering
\smallplotfour{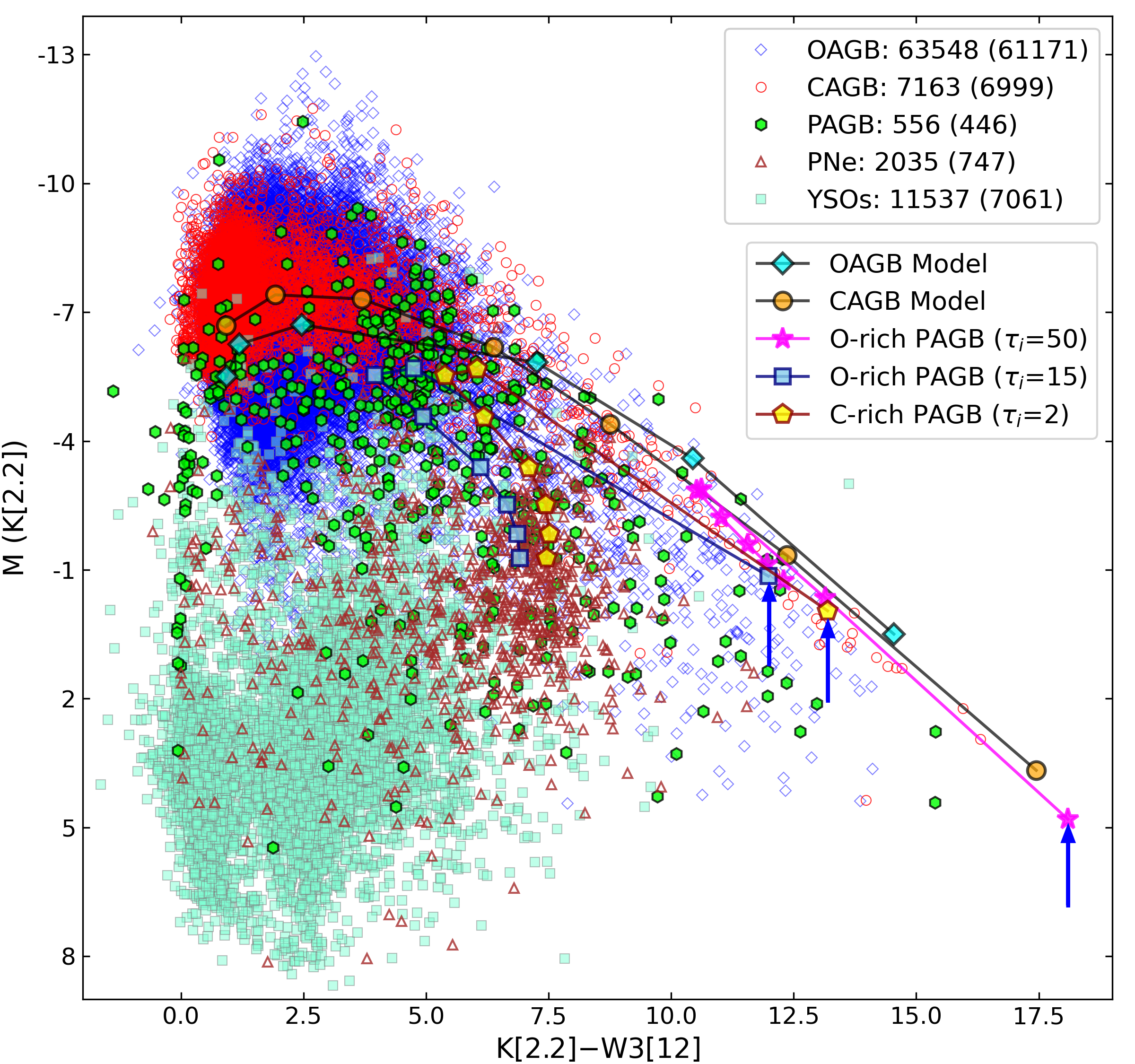}{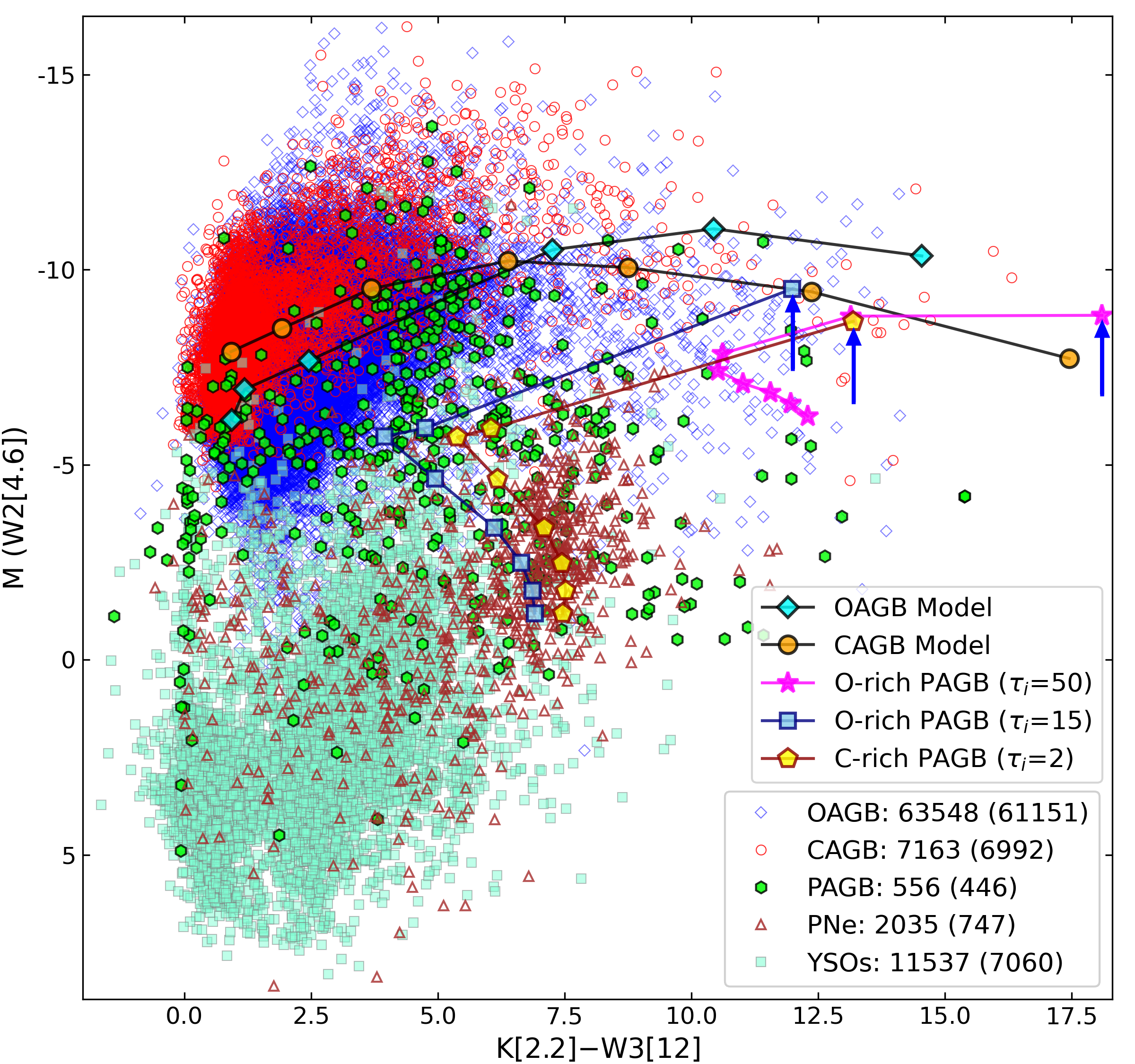}{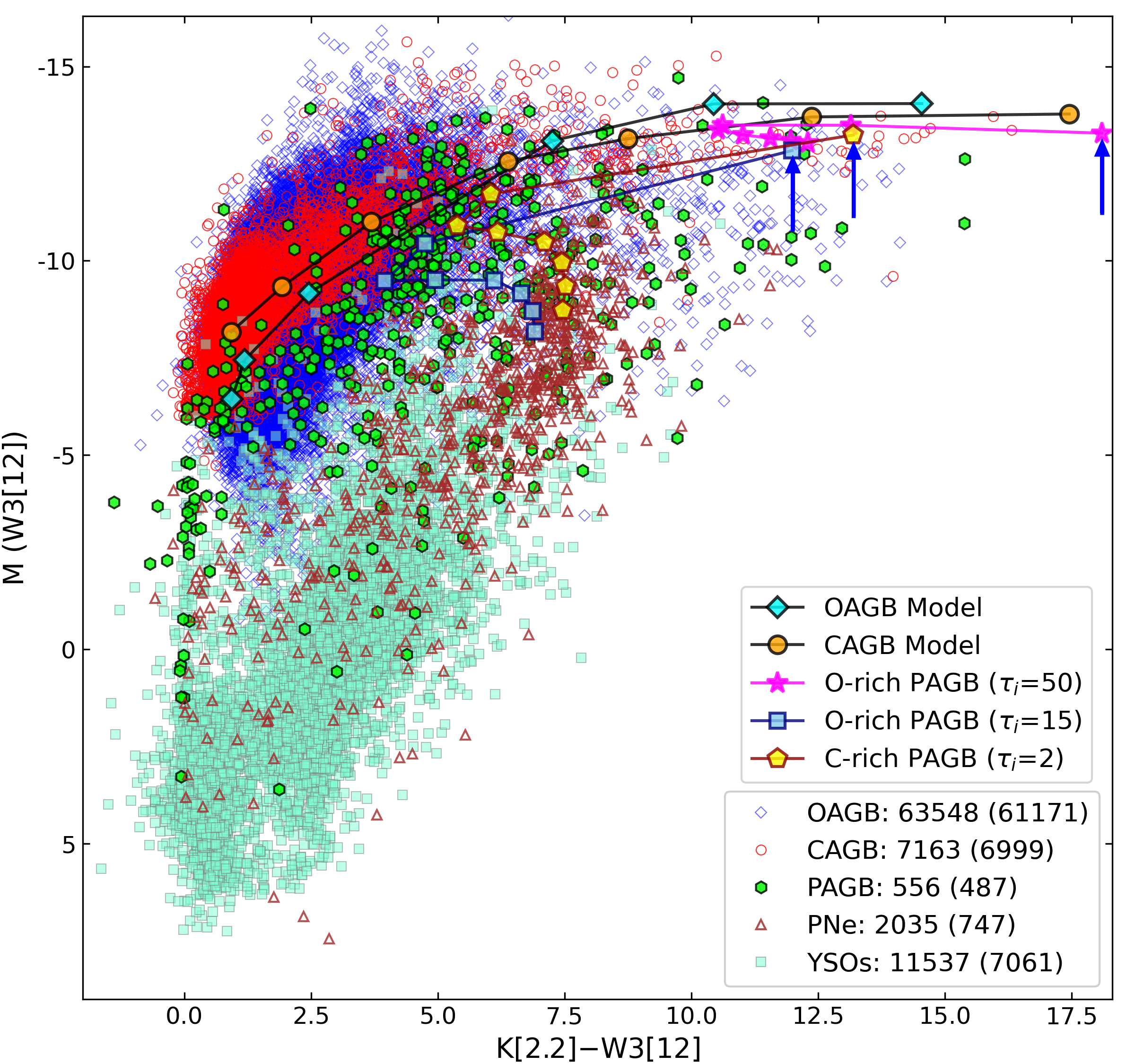}{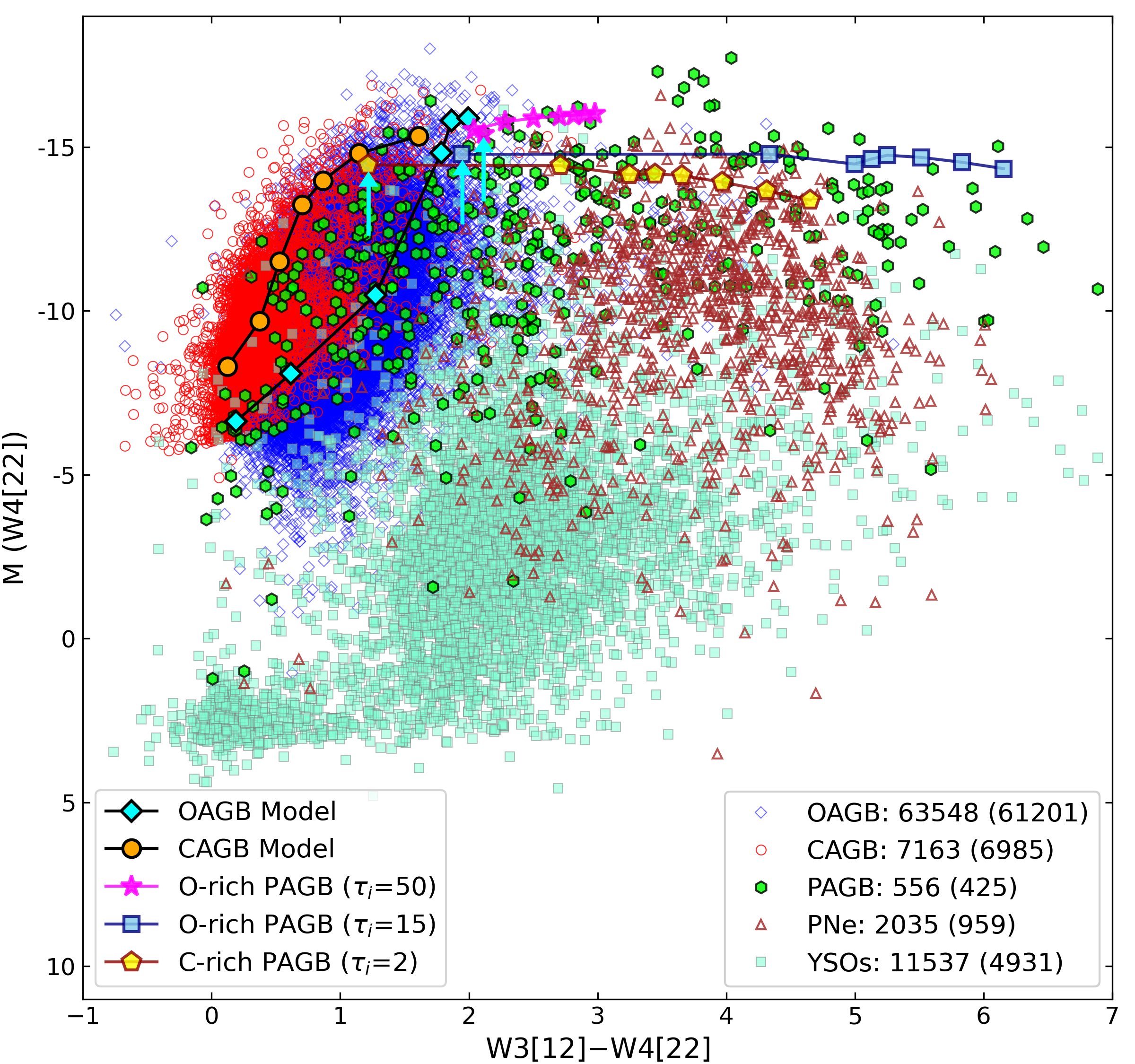}\caption{WISE-2MASS CMDs for the five classes of sample stars (see Table~\ref{tab:tab1}).
For CAGB models: $\tau_{10}$ = 0.0001, 0.01, 0.1, 0.5, 1, 2, and 4 from left to right.
For OAGB models: $\tau_{10}$ = 0.001, 0.01, 0.1, 7, 15, and 30 from left to right (see Section~\ref{sec:agb-model}).
For each class, the number of objects is indicated.
The number in parentheses represents the count of plotted objects with good-quality observational data.
The blue (or cyan) arrows indicate the start points of the PAGB models with various $\tau_{i}$ and
one PML model (1*PML) (see Section~\ref{sec:pagb-model}).
\label{f8}}
\end{figure*}

We plot the theoretical model tracks for PAGB stars on the IR 2CDs and CMDs in
Figures~\ref{f5},~\ref{f6}, and~\ref{f8}. On the IR 2CDs and CMDs, the tip of an
arrow indicates the point at the start of the PAGB phase for each PAGB model
track.

\begin{figure}[t]
\centering
\includegraphics[width=88mm]{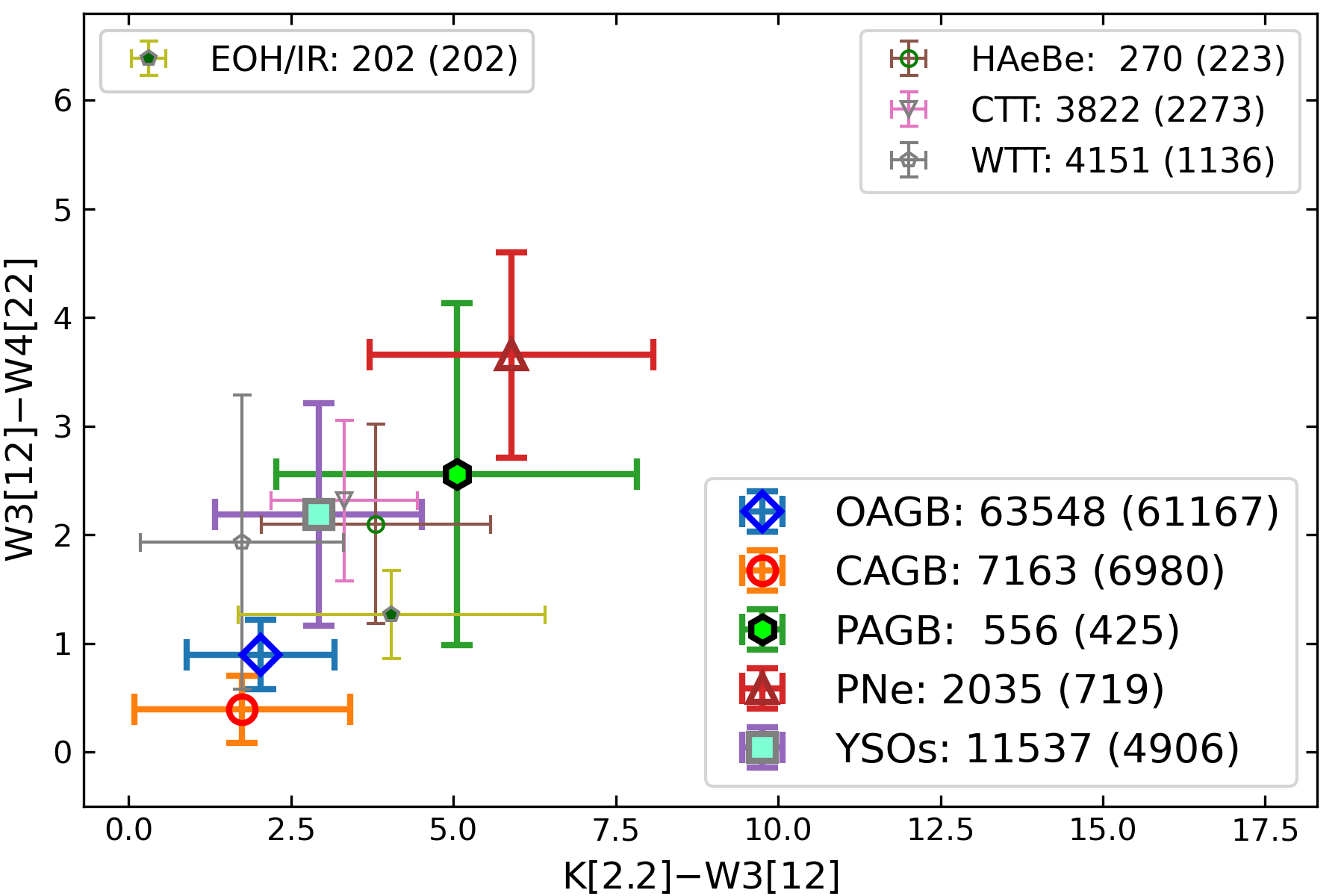}
\caption{An error-bar plot for the WISE-2MASS 2CD for the five classes of sample stars (see Figure~\ref{f6}).
Additionally, plots for three subclasses of YSO and one subclass (EOH/IR) of OAGB are shown.
\label{f7}}
%\vspace{5mm} %% add extra space ONLY when figures/tables are "colliding"!
\end{figure}

\subsection{Limitations of the theoretical models\label{sec:limit}}

Although the theoretical dust shell models used in this work can be useful for
reproducing various observations of AGB and PAGB stars in IR bands, these models
do not account for gas-phase radiation processes or non-spherical dust envelopes.

AGB and PAGB stars showcase a variety of gas-phase emission or absorption
features in near-infrared (NIR) and visual bands (e.g., \citealt{suh2020}).
Consequently, differences between theoretical models and observations are likely
to be more noticeable in wavelength bands where gas-phase radiation processes are
prominent. Additionally, for AGB and PAGB stars with non-spherical dust
envelopes, observed colors and/or magnitudes may demonstrate notable deviations
from theoretical models in both visual and IR bands (e.g., \citealt{suh2016}).

\begin{figure*}
\centering
\xsmallplotfour{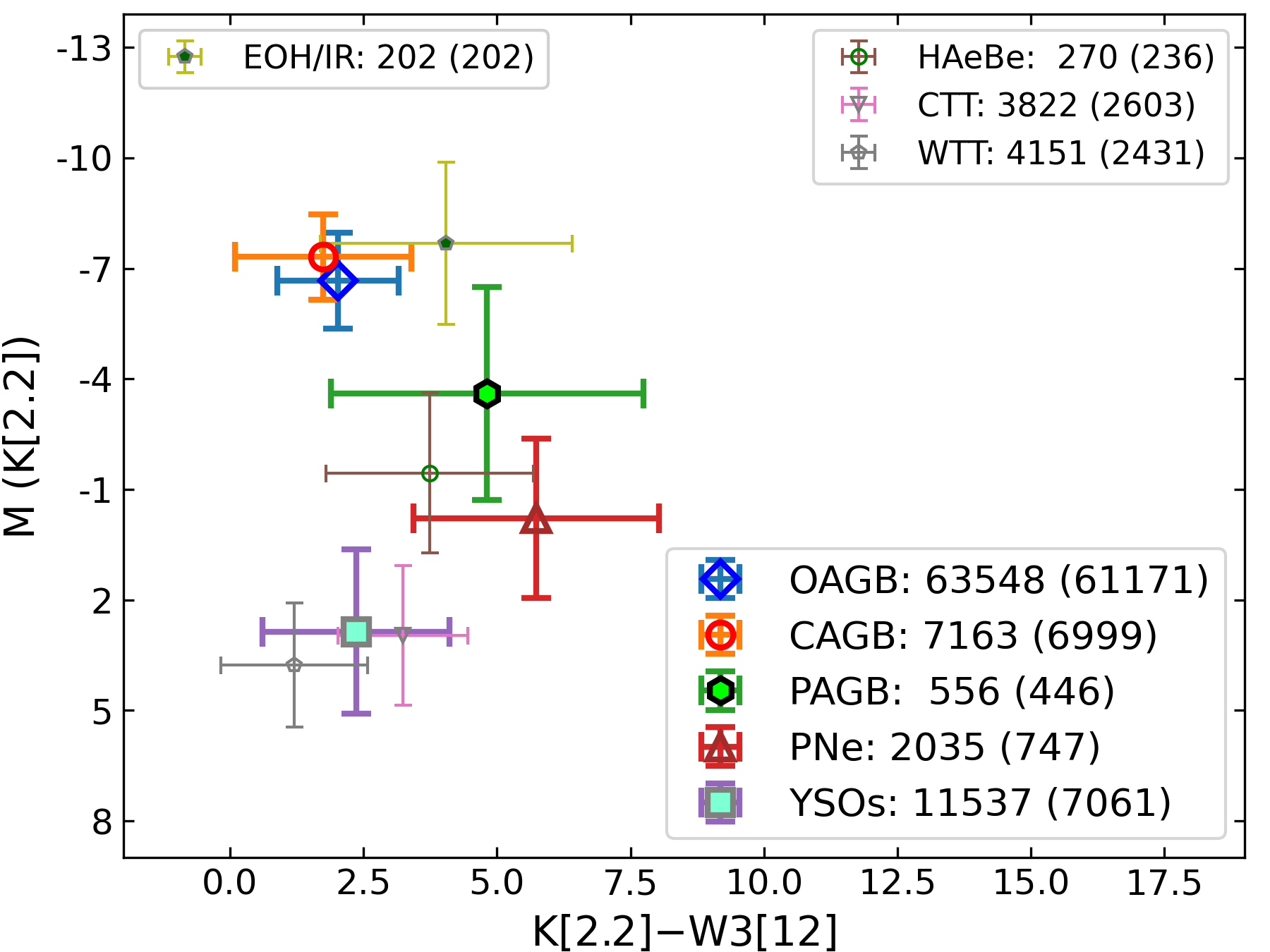}{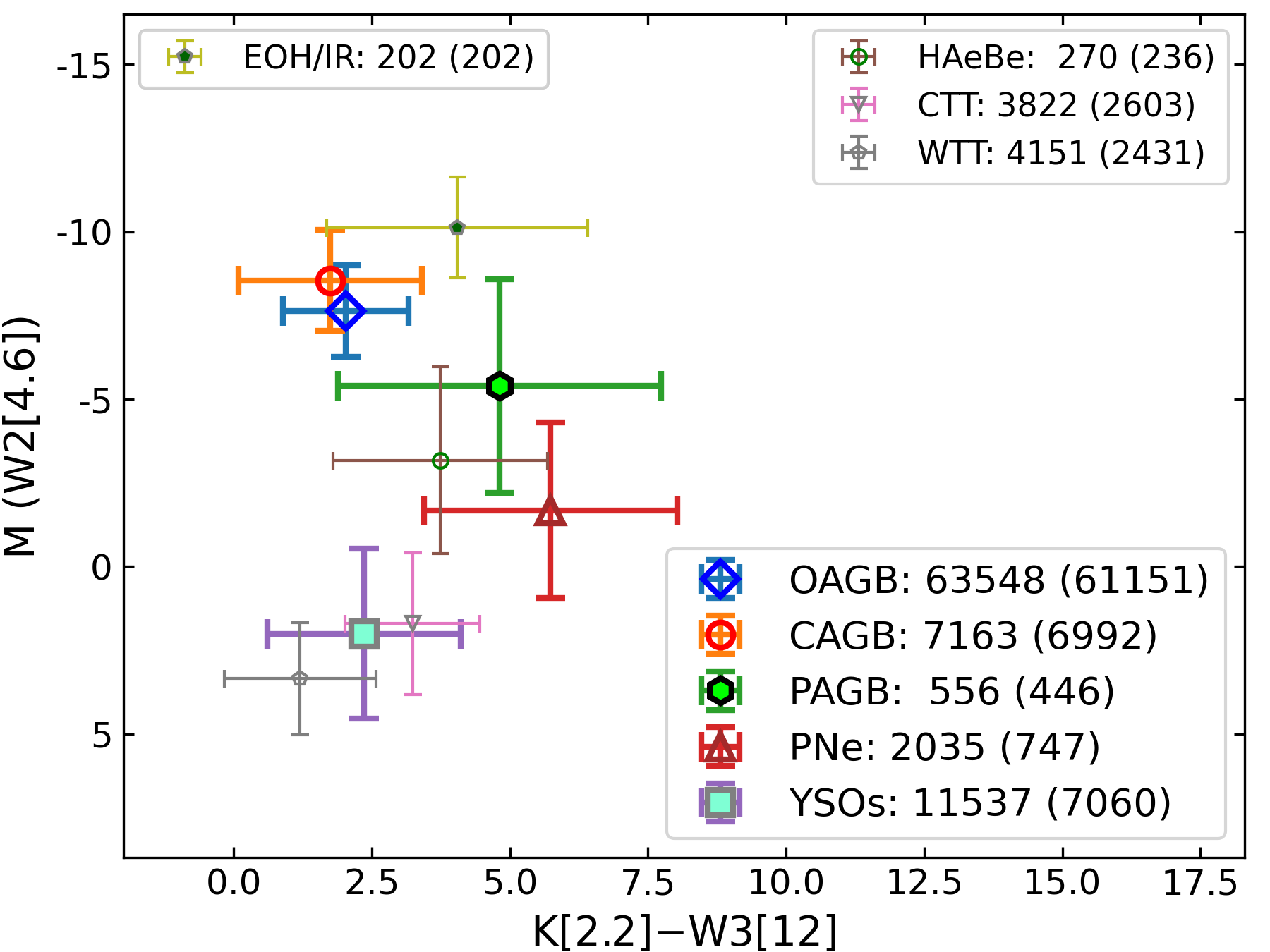}{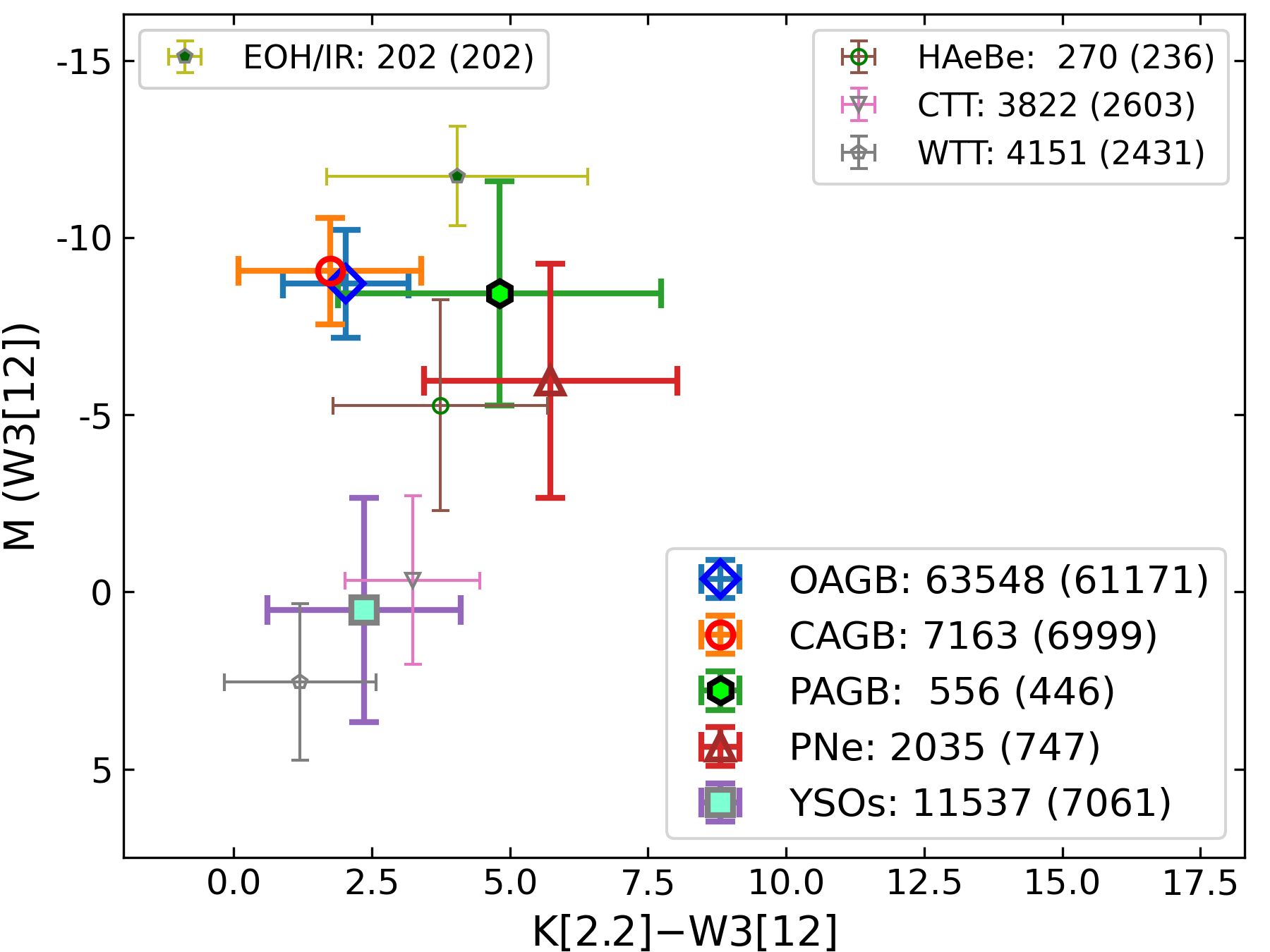}{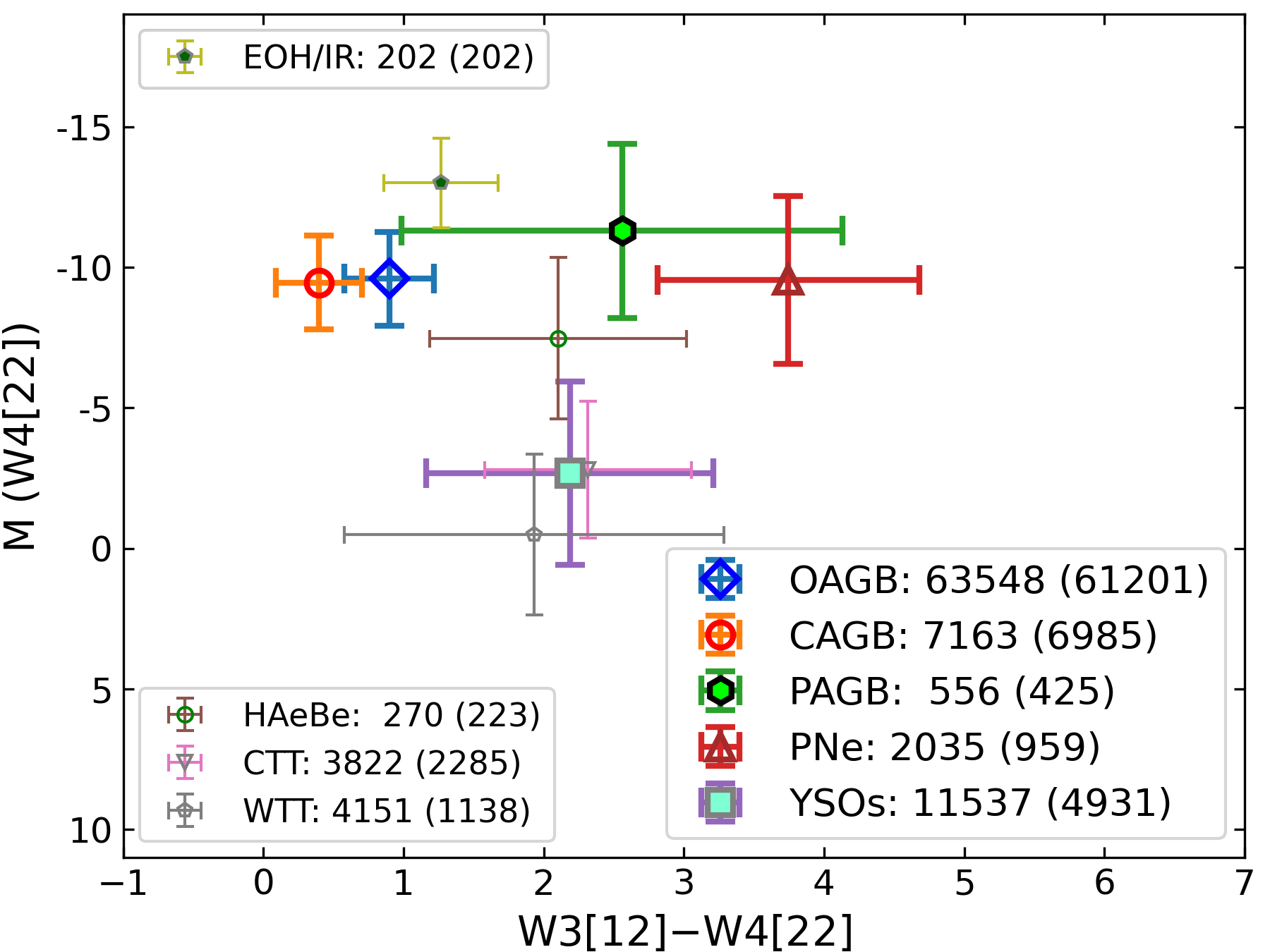}\caption{
Error bar plots for the WISE-2MASS CMDs for the five classes of sample stars (see Figure~\ref{f8}).
Additionally, plots for three subclasses of YSO and one subclass (EOH/IR) of OAGB are shown.
\label{f9}}
\end{figure*}

\section{Properties in IR bands\label{sec:irpro}}

In this section, we explore the IR properties of five classes of sample stars
(see Section~\ref{sec:sample}). Utilizing various observational datasets, we
present IR 2CDs and IR CMDs for the sample stars, alongside theoretical models
for AGB and PAGB stars (refer to Section~\ref{sec:models}), and describe the
characteristics for each class of sample stars. The IR bands utilized for the IR
2CDs and CMDs presented in this study are listed in Table~\ref{tab:tab2}.

On all IR 2CDs and CMDs in Figures~\ref{f5}, \ref{f6}, and \ref{f8}, theoretical 
models for AGB and PAGB stars (see Section~\ref{sec:models}) are plotted for 
comparison with the observations.

\begin{figure*}
\centering
\xsmallplotsix{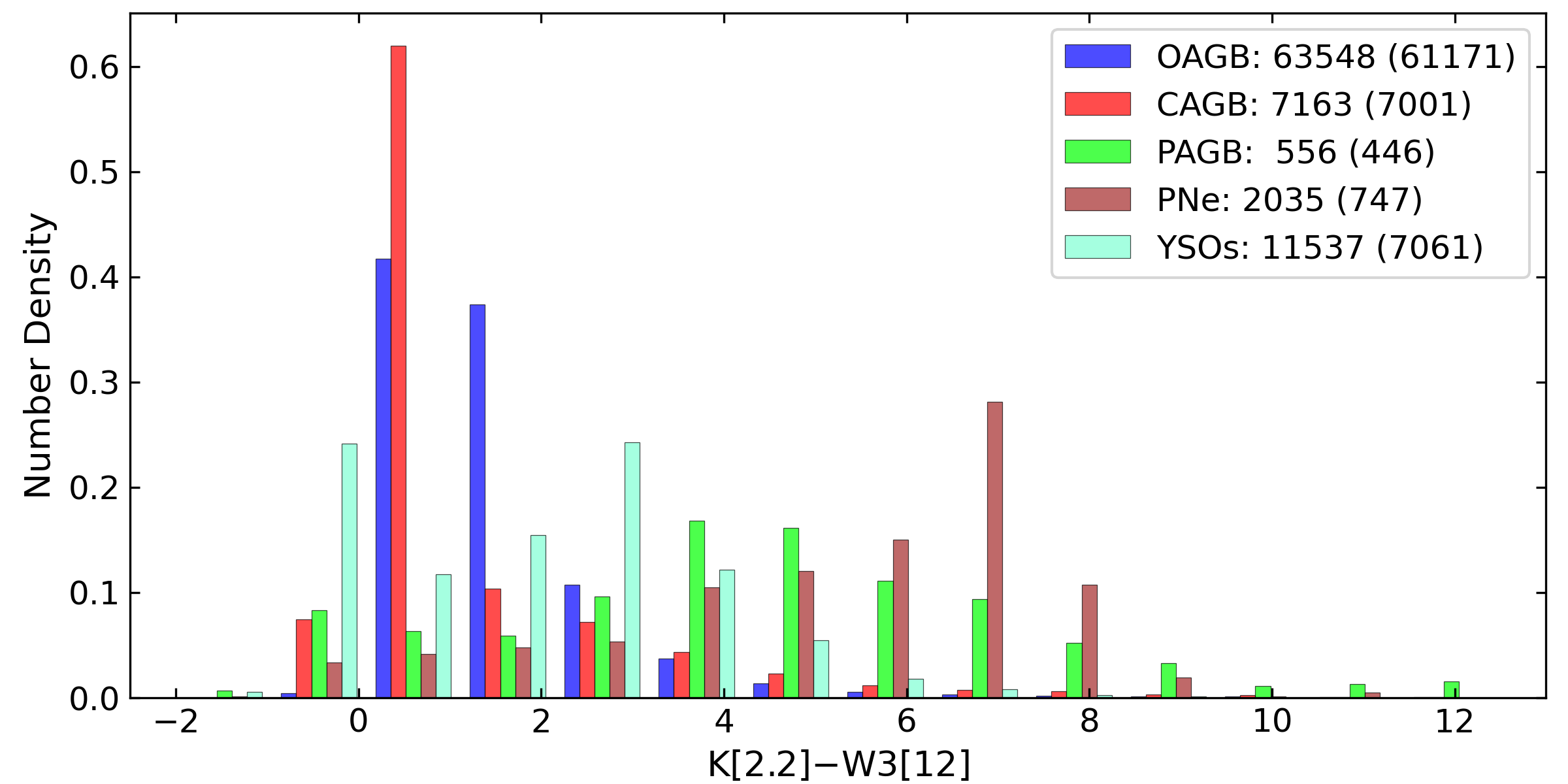}{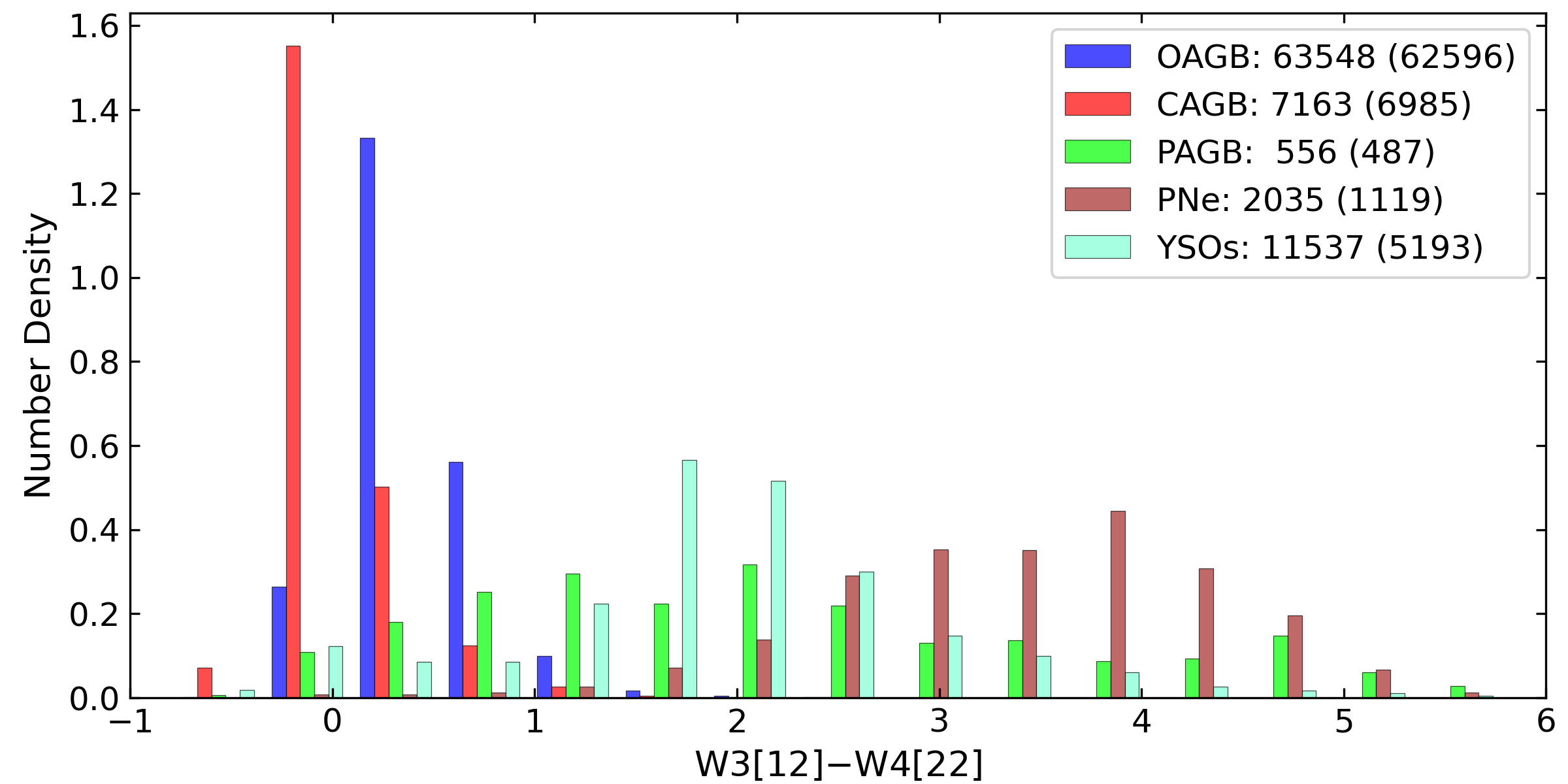}{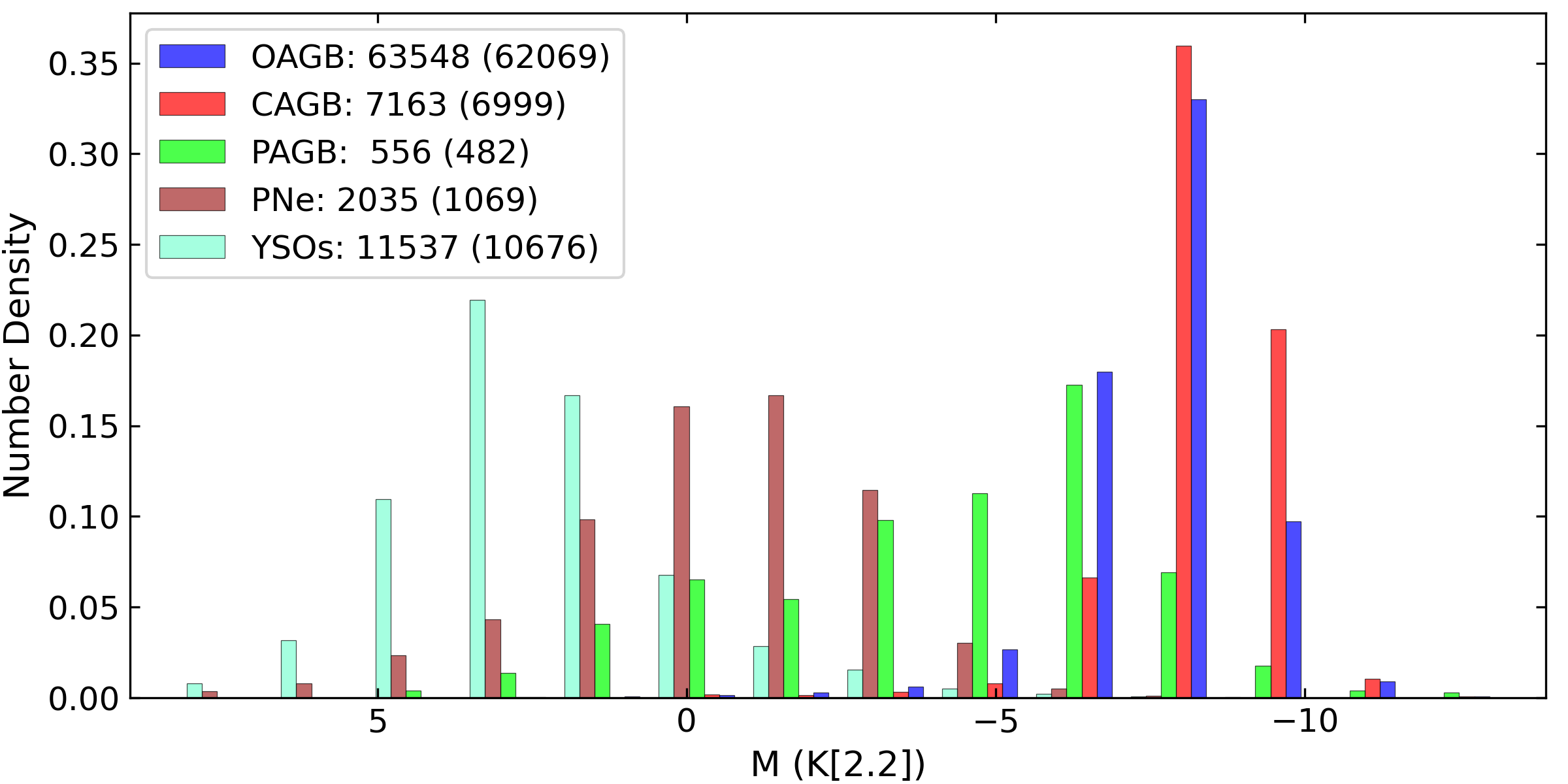}{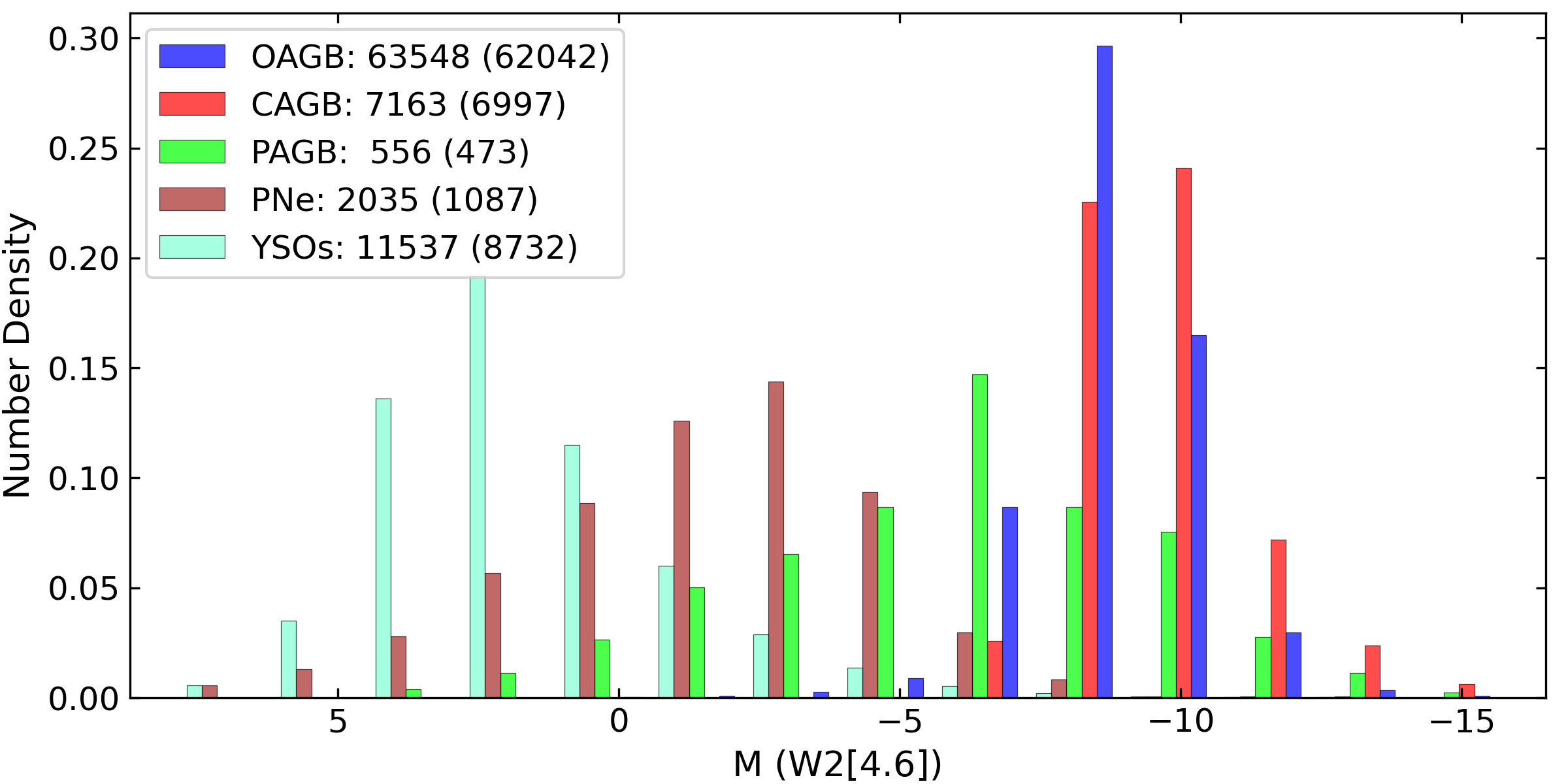}{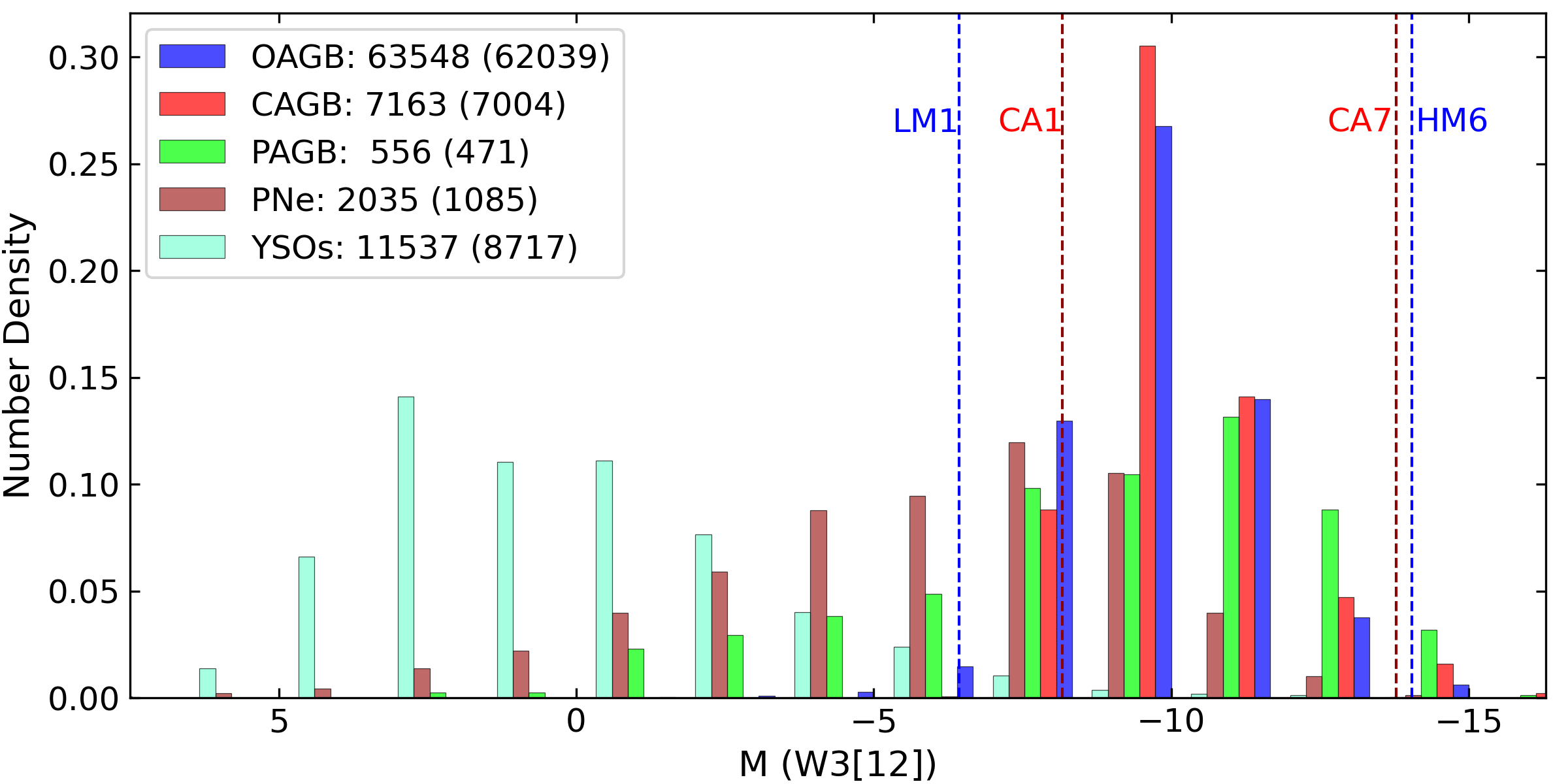}{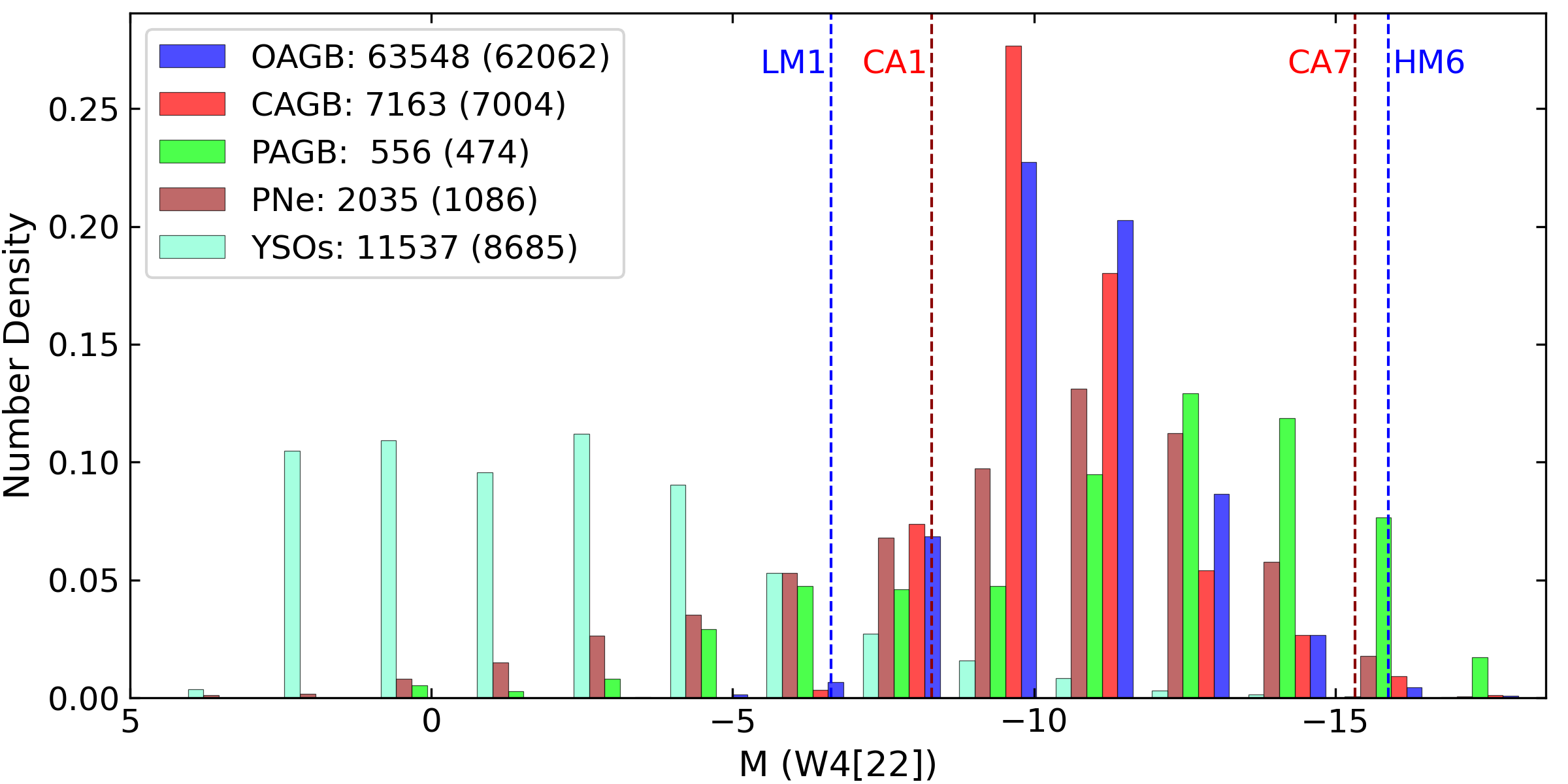}\caption{
Histograms of observed IR colors and absolute magnitudes for the five classes of sample stars.
The vertical lines in lower two panels indicate theoretical magnitudes for AGB models (see Table~\ref{tab:tab3}).
\label{f10}}
\end{figure*}

\subsection{IR 2CDs\label{sec:ir2cd}}

The IR 2CDs have proven valuable for investigating the chemical and physical 
attributes of diverse celestial objects (e.g., \citealt{vanderveen1988}; 
\citealt{suh2015}). Figures~\ref{f5} and~\ref{f6} show IR 2CDs utilizing 
different combinations of observed IR colors for various classes of sample stars 
(refer to Tables~\ref{tab:tab1}). 

The upper panel of Figure~\ref{f5} illustrates an IRAS 2CD plot depicting 
IR[25]$-$IR[60] versus IR[12]$-$IR[25] for the sample stars with IRAS 
counterparts. We observe that the theoretical models for AGB and PAGB stars 
provide an initial explanation for the distribution of observed data points. In 
this study, we divide this 2CD into four distinct regions (R1 to R4) to refine 
the classification process. Most AGB stars are situated in regions R1 and R2, 
whereas most PNe and PAGB stars occupy region R3. 

We find that PAGB models converge in R3, where most PNe and PAGB stars are 
located. Notably, PNe are more densely clustered in R3, while PAGB stars exhibit 
a more dispersed distribution across other regions. PAGB stars and PNe in R4 are 
likely to have non-spherical dust envelopes or be binary systems (e.g., 
\citealt{suh2015}). 

On the IRAS 2CD plot, CAGB stars form a 'C'-shaped trajectory, predominantly
positioned in the lower-left region (R1) and extending towards the right side
(R2). A subset of stars in the upper-left region (R4) consists of visual CAGB
stars (see \citealt{suh2024}).

The lower panel of Figure~\ref{f5} shows an IRAS-2MASS 2CD plot using 
IR[12]$-$IR[25] versus K[2.2]$-$IR[12] for the sample stars with IRAS 
counterparts. Galactic extinction is considered for the K[2.2]$-$IR[12] color. We 
find that OAGB and CAGB stars are more distinguishable on this 2CD. 

Figure~\ref{f6} presents a WISE-2MASS 2CD utilizing W3[12]$-$W4[22] versus 
K[2.2]$-$W3[12] for all sample stars. Galactic extinction is considered for the 
K[2.2]$-$W3[12] color. Overall, the theoretical dust shell models for both OAGB 
and CAGB stars exhibit effective performance in replicating the observed points 
on this IR 2CD. Figure~\ref{f7} displays an error-bar plot for the averaged IR 
colors for the 2CD. 

Significant distinctions arise between PNe and AGB stars, with PNe, characterized 
by more detached dust envelopes, displaying redder W3[12]$-$W4[22] colors. 
Theoretical dust shell model tracks for PAGB stars can roughly emulate the 
general evolutionary path from AGB stars to PNe, showing increasingly redder 
W3[12]$-$W4[22] colors toward the end of the PAGB phase. 

YSOs are positioned in the upper-left and middle-left regions of the WISE-2MASS
2CD, overlapping with the locations of OAGB and PAGB stars.

\subsection{IR CMDs\label{sec:ir-cmd}}

The obtained distance (\citealt{bailer-jones2021}) and extinction data (e.g.,
\citealt{lallement2022}), derived from Gaia DR3 data, are valuable for
determining absolute magnitudes and unreddened colors across a wide range of the
spectrum (Section~\ref{sec:ext}). Galactic extinction is taken into account when
computing the K[2.2]$-$W3[12] color, as well as the absolute magnitude in the
K[2.2] band.

Figure~\ref{f8} shows IR CMDs using K[2.2]$-$W3[12] and W3[12]$-$W4[22] colors 
and absolute magnitudes in the K[2.2], W2[4.6], W3[12], and W4[22] bands for five 
classes of sample stars. The distinction between AGB stars and other classes is 
more pronounced in these IR CMDs. AGB and PAGB stars appear to be the brightest 
among the five classes. 

As we observed for the IR 2CDs discussed in Section~\ref{sec:ir2cd}, the 
theoretical models for AGB and PAGB stars provide a initial explanation for the 
distribution of observed data points on the IR CMDs. We find that PAGB models 
converge in the region where most PAGB stars and PNe are located. 

Because objects with thicker dust shells exhibit redder K[2.2]$-$W3[12] colors,
more evolved AGB stars tend to display a redder K[2.2]$-$W3[12] color. Less
evolved AGB stars with thin dust shells typically show brighter absolute
magnitudes in the K[2.2] band compared to more evolved AGB stars. In contrast,
more evolved (or more massive) AGB stars with thick dust shells are brighter in
the W3[12] and W4[22] bands. The theoretical models for AGB stars also
demonstrate this trend.

Figure~\ref{f9} displays an error-bar plot for the averaged IR colors and
magnitudes for the CMDs. For YSOs, the more massive HAeBe stars exhibit
properties similar to those of PAGB stars or PNe. The less massive CTT and WTT
stars are located in the lower region (see Figure~\ref{f9}). Generally, HAeBe
stars are brighter than CTT stars, and CTT stars are brighter than WTT stars
across all wavelength bands.

We observe that the IR CMDs, incorporating recent distance and extinction data 
from Gaia DR3 for a majority of sample stars, can offer greater effectiveness 
than IR 2CDs in discerning between different classes in many cases.

\subsection{Overall IR properties\label{sec:oirpro}}

Figure~\ref{f10} shows histograms of K[2.2]$-$W3[12] and W3[12]$-$W4[22] colors
and absolute magnitudes at K[2.2], W2[4.6], W3[12], and W4[22] bands for various
classes of sample stars.

We observe that AGB stars are the brightest class in the K[2.2], W2[4.6], and 
W3[12] bands. However, in the W4[22] band, PAGB stars and PNe emit as strongly as 
AGB stars due to the intense emission from their detached dust shells, which 
contributes significantly to the emission in the W4[22] band (refer to the PAGB 
model tracks in Figure~\ref{f8}). Additionally, EOH/IR stars, a subclass of OAGB 
stars, exhibit the highest brightness across all IR bands (see Figure~\ref{f9}). 

In contrast, most YSOs are notably distinguishable from AGB stars on various IR 
CMDs, displaying dimmer absolute magnitudes in IR bands. HAeBe stars, the 
brightest subclass within the YSO class, show comparable brightness to PNe in IR 
bands. 

For W3[12]$-$W4[22] colors, PNe and PAGB stars exhibit systematically redder 
colors compared to AGB stars. This is because PNe and PAGB stars have more 
detached dust envelopes, resulting in redder W3[12]$-$W4[22] colors (see the PAGB 
model tracks in the last panel of Figure~\ref{f8}). Generally, PNe have more 
detached dust envelopes than PAGB stars, and PAGB stars have more detached dust 
envelopes than AGB stars. Therefore, as AGB stars evolve into PAGB stars and PNe, 
their W3[12]$-$W4[22] colors become redder due to the increasing detachment of 
their dust shells. 

The lower two panels of Figure~\ref{f10} display histograms of absolute 
magnitudes in the W3[12] and W4[22] bands for the five classes of sample stars, 
alongside the theoretical model colors of AGB stars (refer to 
Table~\ref{tab:tab3}). The magnitude distribution of AGB stars is more confined 
to narrow regions, closely matching the model colors of AGB stars, whereas the 
distribution for YSOs is spread across much wider regions. This is because the 
absolute magnitudes of YSOs vary significantly depending on the subclass (see 
Figure~\ref{f9}).

\subsection{YSOs and AGB stars\label{sec:YSO-AGB}}

\citet{robitaille2008}, \citet{gutermuth2009}, and \citet{koenig2014} utilized 
various IR 2CDs and CMDs to differentiate YSOs from AGB stars. They used similar 
IR 2CDs, yielding results consistent with those of this study. However, 
\citet{koenig2014} employed IR CMDs using apparent magnitudes, which are not 
effective for comparing YSOs and AGB stars in our Galaxy on a large scale. 
Additionally, they did not present proper theoretical models for AGB and PAGB 
stars compared with the observations. 

In this study, we have presented IR CMDs of YSOs and AGB stars that are 
significantly more useful, thanks to the use of absolute magnitudes derived from 
Gaia DR3 distances. Our sample star selection is more extensive, and the 
observations (IR 2CDs and IR CMDs) are compared with theoretical models for AGB 
and PAGB stars. Generally, IR CMDs using absolute magnitudes are more effective 
than IR 2CDs in distinguishing YSOs from AGB stars in our Galaxy (see 
Sections~\ref{sec:ir2cd}, \ref{sec:ir-cmd}, and \ref{sec:oirpro}).

\section{Summary\label{sec:sum}}

We have explored properties of OAGB stars, CAGB stars, PAGB stars, PNe, and YSOs 
in our Galaxy through an analysis of observational data across visual and IR 
bands. Leveraging datasets from IRAS, 2MASS, AllWISE, Gaia DR3, and SIMBAD, we 
have undertaken a comprehensive comparison between observational data and 
theoretical models. This comparison entails the utilization of diverse CMDs and 
2CDs. 

Utilizing the recently obtained distance and extinction data derived from Gaia 
DR3, we have streamlined the process of determining absolute magnitudes and 
unreddened colors across a wide spectrum. Leveraging this information for a 
majority of sample stars in our Galaxy, we have presented a variety of CMDs and 
2CDs that prove valuable for discerning between different classes within the 
sample stars. 

We presented diverse IR 2CDs and CMDs for distinct classes of sample stars, 
juxtaposed with theoretical models of AGB and PAGB stars. This comparison aims to 
unveil potential differences in their IR properties. We have conducted radiative 
transfer model computations for AGB and PAGB stars, examining diverse parameters 
of central stars and spherically symmetric dust shells. A thorough comparison of 
the theoretical models with observations across various IR 2CDs and CMDs shows a 
significant agreement. 

We observed that PNe exhibit redder W3[12]$-$W4[22] colors than PAGB stars, and 
PAGB stars display redder colors than AGB stars. This trend occurs because as AGB 
stars evolve into PAGB stars and PNe, their W3[12]$-$W4[22] colors become redder 
due to the increasing detachment of their dust shells.

We have found that the CMDs, which incorporate distance and extinction data from 
Gaia DR3 for a majority of sample stars, can be more effective than 2CDs in 
distinguishing between different classes in many cases. 

In visual bands, AGB and PAGB stars are among the brightest classes. Generally, 
AGB and PAGB stars are brighter than PNe, and PNe are brighter than YSOs in 
visual bands. However, HAeBe stars, a subclass of YSOs, show comparable 
brightness to PAGB stars in visual bands. 

In IR bands, EOH/IR stars, a subclass of OAGB stars, are the brightest. AGB stars 
are the brightest class in K[2.2], W2[4.6], and W3[12] bands. However, in the 
W4[22] band, PAGB stars and PNe are as bright as AGB stars due to the stronger 
emission produced by their detached dust shells. In contrast, most YSOs are 
notably distinguishable from AGB stars on various IR CMDs, displaying dimmer 
absolute magnitudes in IR bands. HAeBe stars, the brightest subclass within the 
YSO class, show comparable brightness to PNe in IR bands. 

For EOH/IR stars, which are the brightest subclass among the AGB classes in IR 
bands, the absolute magnitudes in visual bands can be dimmer than those of other 
subclasses. This is because the thick dust envelopes surrounding EOH/IR stars 
cause greater extinction of emission in visual bands. For YSOs, HAeBe stars are 
brighter than CTT stars, and CTT stars are brighter than WTT stars across all 
wavelength bands.

%%% ACKNOWLEDGMENTS (IF ANY) %%%%%%%%%%%%%%%%%%%%%%%%%%%%%%%%%%%%%%%%

\acknowledgments

I thank the anonymous referee for constructive comments and suggestions. This 
research was supported by Basic Science Research Program through the National 
Research Foundation of Korea (NRF) funded by the Ministry of Education 
(2022R1I1A3055131). This research has made use of the SIMBAD database and VizieR 
catalogue access tool, operated at CDS, Strasbourg, France. This research has 
made use of the NASA/IPAC Infrared Science Archive, which is operated by the Jet 
Propulsion Laboratory, California Institute of Technology, under contract with 
the National Aeronautics and Space Administration.

%%% APPENDICES (IF ANY) %%%%%%%%%%%%%%%%%%%%%%%%%%%%%%%%%%%%%%%%%%%%%

%%% LIST OF REFERENCES (natbib STYLE) %%%%%%%%%%%%%%%%%%%%%%%%%%%%%%%

%%% Option 1: Use \bibliography %%%%%%%%%%%%%%%%%%%%%%%%%%%%%%%%%%%%%
% \bibliography{jkas-sample}

%%% Option 2: Use \begin{thebibliography} ... \end{thebibliography}%%

%%% END LIST OF REFERENCES %%%%%%%%%%%%%%%%%%%%%%%%%%%%%%%%%%%%%%%%%%

\end{document}